\documentclass[a4,12pt]{article}
\newcommand{\nn}{\nonumber\\}

\newcommand{\ket}[1]{\left| #1 \right>}
\newcommand{\abs}[1]{\left| #1 \right|}

\newcommand{\Q}{Q_{\rm B}}

\newcommand{\Bra}{\Big< \hspace{-1mm}\Big<}
\newcommand{\Ket}{\Big> \hspace{-1mm}\Big>}
\renewcommand{\thepage}{}
\makeatletter
\@addtoreset{equation}{section}
\renewcommand{\theequation}{\thesection.\@arabic\c@equation}
\makeatother
\renewcommand{\thefootnote}{\fnsymbol{footnote}}
\begin{document}
\begin{titlepage}
\title{
\vspace*{-4ex}
\hfill
\begin{minipage}{3.5cm}
\normalsize KEK-TH-1024\\
\normalsize hep-th/0506240
\end{minipage}\\
\vspace{4ex}
\bf Marginal Deformations and Classical Solutions in Open Superstring
 Field Theory 
\vspace{5ex}}
\author{Isao {\sc Kishimoto}$^{1,}$\footnote{E-mail address:
ikishimo@post.kek.jp}\ \ and\ \  
Tomohiko {\sc Takahashi}$^{2,}$\footnote{E-mail address:
tomo@asuka.phys.nara-wu.ac.jp}\\
\vspace{2ex}\\
$^1${\it High Energy Accelerator Research Organization (KEK),}\\
{\it Tsukuba, Ibaraki 305-0801, Japan}\\
$^2${\it Department of Physics, Nara Women's University,}\\
{\it Nara 630-8506, Japan}}
\date{June, 2005}
\maketitle
\vspace{7ex}

\begin{abstract}
\normalsize
\baselineskip=19pt plus 0.2pt minus 0.1pt

We construct a class of classical solutions in the
Berkovits' open superstring field theory. The resulting solutions
correspond to marginal boundary deformations in conformal field theory. 
The vacuum energy vanishes exactly for the solutions.
Investigating the theory expanded around one of the solutions, we find
that it reflects the effect of background Wilson lines.
The solution has a well-defined Fock space expression and it is
invariant under space-time supersymmetry transformation.
\end{abstract}
\end{titlepage}

%%%%%%%%%%%%%%%%%%%%%%%%%%%%%%%%%%%%%%%%%%%%%%%%%%%%%%
\renewcommand{\thepage}{\arabic{page}}
\renewcommand{\thefootnote}{\arabic{footnote}}
\setcounter{page}{1}
\setcounter{footnote}{0}
\baselineskip=19pt plus 0.2pt minus 0.1pt
%%%%%%%%%%%%%%%%%%%%%%%%%%%%%%%%%%%%%%%%%%%%%%%%%%%%%%
%
%%%%%%%%%%%%%%%%%%%%%%%%%%%%%%%%%%%%%%%%%%%%%%%%%%%%%%
\section{Introduction}

String field theory is established as a framework for exploring
nonperturbative structures in string theory. 
Motivated by Sen's conjecture \cite{rf:Sen1,rf:Sen2}, many people
studied classical solutions extensively and intriguing
results were provided in bosonic open string field theory
\cite{rf:SZ,rf:Senreview}.
In the supersymmetric case, the most promising theory is formulated in
terms of
the Wess-Zumino-Witten (WZW) like action proposed by Berkovits
\cite{rf:SSFT1,rf:SSFT2}, which has no problem with contact term
divergences \cite{rf:wendt}.   
This open superstring field theory is a sufficient framework to
elucidate nonperturbative phenomena of the Neveu-Schwarz (NS) sector.
Indeed, the tachyon vacuum and kink solutions were found in the
superstring field theory by using the level truncation
scheme \cite{rf:SSFT2,rf:BSZ,rf:DSR,rf:IN,rf:Ohmori}. On the
analytical side,  there are some attempts to construct exact solutions
in terms of a half string formulation \cite{rf:Kluson}, a pregeometrical
formulation \cite{rf:Kluson2,rf:Sakaguchi}, a conjecture of
vacuum superstring field theory
\cite{Arefeva:2001ke,rf:vssft1,rf:vvsft2,Arefeva:2002sg,rf:vvsft3}
and an analogy with integrable systems \cite{rf:LPU}. 

In the present paper we construct analytic classical solutions  
in the open superstring field theory using techniques developed in
bosonic open string field theory~\cite{rf:TT1,rf:TT2,rf:KTZ}. 
The resulting solution consists of the identity string field, ghost
fields and an operator associated with a current. Taking $\partial X(z)$
as the current, we find that the action expanded around the
solution can be transformed to the original action by a string field
redefinition. In the redefined theory, however, the momentum
is shifted in the string field and then the classical solution can be
related to a background Wilson line. 
Generically, we anticipate that our solutions correspond to marginal
boundary deformed backgrounds as in the bosonic case.  

The analytic solutions are useful for studying gauge
structure in string field theory.
In bosonic string field theory, the analytic solution
corresponding to Wilson lines can be represented as a
``locally'' pure gauge, 
and then we find that a ``locally'' pure gauge configuration in string
field theory corresponds to a marginal deformation in conformal field
theory~\cite{rf:TT1,rf:TT2,rf:KTZ}. This correspondence is a natural
generalization of that of low energy effective theories. Later we will
see that the solution in the superstring field theory shares this
feature of the bosonic theory. 

Marginal deformations in string field theory were often studied using the
level truncation scheme. We see that the effective potential for a
marginal field becomes flatter as the truncation level is
increased~\cite{rf:LMD,rf:EA-WT,rf:MD-IN,rf:CST,rf:YZ,rf:YZ2}, and then
the vacuum energy of the analytic solution must vanish.
Unfortunately, we encounter a difficulty in calculating
the vacuum energy in the bosonic theory. Though the
vacuum energy formally 
vanishes, it is given as a kind of indefinite quantities if we
calculate it  by oscillator
representation~\cite{rf:TT1,rf:TT2,rf:KTZ}. However, we will see 
that the vacuum energy is to be exactly zero in the superstring
field theory. This result is a characteristic feature of the
supersymmetric case. 

In string field theory, the gauge symmetry includes global
symmetries generated by $K_n=L_n-(-1)^nL_{-n}$
\cite{rf:SCSFT,rf:IIKTZ}. It 
is a typical symmetry in string field theory because the symmetry
mixes various component fields and it has a non-local structure.
Based on an analytical approach, we find that the Wilson line parameter in
the solution is invariant under the global transformation.

Although it is hard to include the Ramond (R) sector
into the action, we have the equations of motion for both of the NS and
R sectors~\cite{rf:Rsector}. The equations of motion possess a fermionic
symmetry which transforms the NS boson (R fermion) to the R fermion (NS
boson). Then we expect that 
the superstring field theory has space-time supersymmetry. 
Actually, we find that global space-time supersymmetry is realized
on-shell as a part of the fermionic symmetry. We show that 
the solution corresponding to a Wilson line is a supersymmetric
solution, namely the solution is invariant under the global space-time
supersymmetry transformation. 

This paper is organized as follows. In section 2 we construct an
analytic classical solution in the open superstring field
theory. We find that the solution can be written by a well-defined Fock
space expression. After discussing the vacuum energy and the theory
expanded around the 
solution, we relate the solution to background Wilson lines.
Michishita gives a covariant action of the R sector by imposing a
constraint equation \cite{rf:Michi}.
We discuss the effect of the
solution on the R sector in terms of the action proposed by Michishita.
Moreover, we
show how the solution is transformed under the global
symmetry and space-time supersymmetry.
In section 3 we extend the Wilson line solution to those which
correspond to general marginal boundary deformations generated by
supercurrents. 
We find that generalized solutions also have a favorite feature that the
vacuum energy vanishes.
We offer some comments related to our results and discuss open questions
in section 4.  
In addition we include four appendices. We
represent the identity string field by explicit oscillator expression in
the large Hilbert space in appendix A. In appendix B we give a different
derivation of the action expanded around a general solution which is
originally given in refs.~\cite{rf:vssft1,rf:Kluson}. We use an
alternative expression of the action given in ref.~\cite{rf:BOZ} to
derive the expanded action.
In appendix C we show that the fermionic symmetry contains
global space-time supersymmetry, and give some comments on
supersymmetry in the cubic superstring field theory \cite{rf:SCSFT} and
its modified theory \cite{rf:PTY,rf:Arefeva}.
In appendix D we construct the analytic solution in bosonic
string field theory which corresponds to a general marginal
boundary deformations.

%%%%%%%%%%%%%%%%%%%%%%%%%%%%%%%%%%%%%%%%%%%%%%%%%%%%%%
\section{Classical solutions and background Wilson lines}

The open superstring field theory action \cite{rf:SSFT1,rf:SSFT2}
is given by 
\begin{eqnarray}
\label{Eq:action}
 S[\Phi]=\frac{1}{2g^2} \Bra\,
(e^{-\Phi} \Q e^\Phi)(e^{-\Phi}\eta_0 e^\Phi)
-\int_0^1 dt\,(e^{-t\Phi}\partial_t e^{t\Phi})
\left\{(e^{-t\Phi}\Q e^{t\Phi}),\,
(e^{-t\Phi}\eta_0 e^{t\Phi})\right\}
\Ket,
\end{eqnarray}
where $\Phi$ denotes a string field of the GSO(+) NS
sector which corresponds to a Grassmann even vertex operator of ghost
number 0 and picture number 0 in the conformal field theory. CFT
correlators 
%$\left<\hspace{-1mm}\left< \cdots \right>\hspace{-1mm}\right>$
$\langle\!\langle\cdots \rangle\!\rangle$ are
defined in the large Hilbert space and 
$\{A,\,B\}\equiv AB+BA$.\footnote{For details of the definition, see
for instance ref.~\cite{rf:BSZ}.}
The action is invariant under the infinitesimal gauge
transformation,
\begin{eqnarray}
 \delta e^{\Phi}=(\Q\delta\Lambda)*e^\Phi+
e^\Phi*(\eta_0\delta\Lambda'),
\end{eqnarray}
where $\delta\Lambda$ and $\delta\Lambda'$ are infinitesimal
parameters.
Integrating this infinitesimal form, we can obtain the finite gauge
transformation as\footnote{The gauge transformation can be
expressed as $e^{\Phi'}=g*e^\Phi*g'$ where $\Q g=\eta_0 g'=0$, since
each of the operators, $\Q$ and $\eta_0$, has trivial cohomology in the
large Hilbert 
space. Here, $g$ and $g'$ are group elements in the ``stringy gauge
group'' in superstring field theory.}
\begin{eqnarray}
\label{Eq:gaugetrans}
 e^{\Phi'} = e^{\Q\Lambda}*e^\Phi*e^{\eta_0\Lambda'},
\end{eqnarray}
where $\Lambda$ and $\Lambda'$ are finite parameters.
Variating the action (\ref{Eq:action}), we can derive the
equation of motion to be 
\begin{eqnarray}
 \eta_0(e^{-\Phi}\,\Q e^\Phi)=0.
\end{eqnarray}

For simplicity, we mainly consider superstring field theory describing
the dynamics of a D brane
without Chan-Paton degrees of freedom. 
We single out a direction on the world volume of the
brane, writing the string coordinate as
$X(z,\bar{z})=(X(z)+X(\bar{z}))/2$ and its supersymmetric  
partner as $\psi(z)$. Our later analysis can be easily
extended to include Chan-Paton indices.

%%%%%%%%%%%%%%%%%%%%%%%%%%%%%%%%%%%%%%%%%%%%%%%%%%%%%%
\subsection{classical solutions in open superstring field theory}

In this subsection, we will show that one of the classical solutions is
given by 
\begin{eqnarray}
\label{Eq:sol}
 \Phi_0 = -\tilde{V}_L(F)I,
\end{eqnarray}
where $I$ is the identity string field
and the operator $\tilde{V}_L$ is 
defined as\footnote{We note that $e^{q\phi}$ ($q$\,:\,odd) is a fermionic
operator. 
More precisely, we need a cocycle factor to represent statistical
property of the operator.}
\begin{eqnarray}
 \tilde{V}_L(F) =\int_{C_{\rm left}}\frac{dz}{2\pi i} F(z)\,
\tilde{v}(z),\ \ \ 
\tilde{v}(z)=\frac{1}{\sqrt{2}}\,c\,\xi\,e^{-\phi}\psi(z).
\end{eqnarray}
Here, $C_{\rm left}$ denotes a counter-clockwise path along a half
of the unit circle, i.e., $-\pi/2<\sigma<\pi/2$ for $z=e^{i\sigma}$.
$F(z)$ is a function on the unit circle $\abs{z}=1$ satisfying
$F(-1/z)=z^2 F(z)$ \cite{rf:TT2,rf:KT}.\footnote{Under this condition,
$F(z)$ cannot be a non-zero constant.}
We have to impose an additional constraint on $F(z)$ due to 
the reality of the string field as explained in the next subsection (see
also appendix A).

First, we introduce half string operators similar to  $\tilde{V}_L(F)$:
\begin{eqnarray}
 C_L(F) &=& \int_{C_{\rm left}}\frac{dz}{2\pi i}F(z)\,c(z),\\
 V_L(F) &=& \int_{C_{\rm left}}\frac{dz}{2\pi i}F(z)\,v(z),
\end{eqnarray}
where $c(z)$ is the ghost field and $v(z)$ is defined as
\begin{eqnarray}
 v(z) = \left[\Q,\,\tilde{v}(z)\right]
=\frac{i}{2\sqrt{\alpha'}}c\partial X(z)+\frac{1}{\sqrt{2}}
\eta e^\phi \psi(z).
\end{eqnarray}
By definition, the commutation relation
$[\Q,\,\tilde{V}_L(F)]=V_L(F)$ holds.
We also define the operators, $C_R(F)$, $V_R(F)$ and
$\tilde{V}_R(F)$ by replacing the integration path $C_{\rm left}$ 
by $C_{\rm right}$ which rotates counter-clockwise along $\abs{z}=1\ \
({\rm Re}\,z\leq 0)$. 
For these half string operators, we can derive
their (anti-)commutation relations
from similar procedures in refs.~\cite{rf:TT2,rf:KT}.
The operator product expansions (OPEs) among local operators in the
integrand are easily calculated as
\begin{eqnarray}
 v(z)\,\tilde{v}(z') &\sim& \frac{1}{z-z'}\,\frac{1}{2}\,c(z'),\\
 v(z)\,v(z') &\sim& \frac{1}{z-z'}\,\frac{-1}{2}(c\partial c(z')
-\eta \partial \eta e^{2\phi}(z')). 
\end{eqnarray}
Using these OPEs, 
we obtain (anti-)commutation
relations of these local operators on the unit circle, $\abs{z}=1$:
\begin{eqnarray}
\label{Eq:vvtilde}
 \left[v(z),\,\tilde{v}(z')\right]  
&=& \frac{1}{2}\,c(z')\,\delta(z,\,z'),\\
\label{Eq:vv}
\{v(z),\,v(z')\} &=& -\frac{1}{2}\,(c\partial c(z')-\eta \partial \eta 
e^{2\phi}(z'))\,\delta(z,\,z'),
\end{eqnarray}
where the delta function is defined as
$\delta(z,\,z')=\sum_{n=-\infty}^\infty z^{n} 
{z'}^{-n-1}$.\footnote{The delta function satisfies
\begin{eqnarray}
 f(z)=\oint_{\abs{z'}=1} \frac{dz'}{2\pi i}\,f(z')\,\delta(z,\,z'),
\end{eqnarray}
where the function $f(z)$ is square integrable on the unit circle
$\abs{z}=1$ $(f \in L^2)$. 
Moreover, for any $f,\ g \in L^2$, the delta function enjoys
the properties, 
\begin{eqnarray}
&&
\int_{C_{\rm left/right}}{dz\over 2\pi i}\!
\int_{C_{\rm left/right}}{dz'\over 2\pi i}
f(z)g(z')\delta(z,z')=
\int_{C_{\rm left/right}}{dz\over 2\pi i}f(z)g(z),\\
&&
\int_{C_{\rm left}}{dz\over 2\pi i}\!
\int_{C_{\rm right}}{dz'\over 2\pi i}f(z)g(z')\delta(z,z')=0,
\end{eqnarray}
which are necessary for derivation of (\ref{Eq:VLVtildL}) and
(\ref{Eq:VLVL}).
}

We integrate (\ref{Eq:vvtilde}) and (\ref{Eq:vv}) to derive 
(anti-)commutation relations of half string operators:
\begin{eqnarray}
\label{Eq:VLVtildL}
 \left[V_L(F),\,\tilde{V}_L(G)\right]&=&\frac{1}{2}\,C_L(FG),\\
\label{Eq:VLVL}
 \left\{V_L(F),\,V_L(G)\right\}&=&-\frac{1}{2}\,
 \left\{\Q,\,C_L(FG)\right\},
\end{eqnarray}
where we have used $\left\{\Q,\,c(z)\right\}=c\partial c(z)
-\eta \partial \eta e^{2\phi}(z)$ in the latter equation. The similar
relations hold for the right-half string operators, and 
other (anti-)commutation relations become zero.
Here, we emphasize that these equations 
hold for any functions $F(z)$ and $G(z)$ 
defined on the unit circle, because we have only to use the
equal-time (anti-)commutation relations to derive the equations.\footnote{
To derive (\ref{Eq:VLVtildL})
and (\ref{Eq:VLVL}), it is sufficient for $F$ and $G$ to be square
integrable. The condition $F(\pm i)=G(\pm i)=0$ is unnecessary for
these (anti-)commutation relations.} 
Namely, the
functions are not necessarily to be holomorphic, though we express them
as functions of a complex variable.

Next, we consider some properties of the half string operators
associated with the star product and the identity string field.
Suppose that two string fields $A$ and $B$ are defined as
$\ket{A}=A(0)\ket{0}$ and $\ket{B}=B(0)\ket{0}$, where $A(z)$ and
$B(z')$ are conformal fields on the unit discs $\abs{z}\leq 1$ and
$\abs{z'}\leq 1$, respectively.  
The star product $A*B$ is defined in terms of the gluing Riemann surface
by the identification $z z'=-1\ \ (\abs{z}=1,\ \ \ {\rm Re}\,z\leq 0)$ 
\cite{LPP1,LPP2}.
Accordingly, it follows that
\begin{eqnarray}
\label{Eq:psiAB}
 \left(\sigma(z)\,A\right)*B=(-1)^{\abs{\sigma}\abs{A}}
A*\left((z'^2)^h\,\sigma(z') B\right)
\ \ \ (zz'=-1,\ \ \abs{z}=1,\ \ {\rm Re}\,z\leq 0),
\end{eqnarray}
where $\sigma(z)$ is a primary field with dimension $h$, and
$\abs{\alpha}$ 
denotes the statistic index defined to be $0 (1)$ if $\alpha$ is a bosonic
(fermionic) operator. 
Multiplying a function $F(z)$ which satisfies
$F(-1/z)=(z^2)^{1-h} F(z)$\ \footnote{$F(z)$ is the same as the function
$F_+^{(-h+1)}(z)$ in ref.~\cite{rf:TT2}.
We note that our analysis is easily extended to the case of $F_{-}(z)$
in ref.~\cite{rf:TT2}.} to both sides of
(\ref{Eq:psiAB}) and 
integrating it along the path $\abs{z}=1\ \ ({\rm Re}\,z\leq 0)$,
we can find the generic formula \cite{rf:TT2,rf:KT}
\begin{eqnarray}
\label{Eq:LRstar}
 \left(\Sigma_R(F) A\right)*B=-(-1)^{\abs{\sigma}\abs{A}}
A*\left(\Sigma_L(F)B\right),
\end{eqnarray}
where the operator $\Sigma_{L(R)}$ is defined as
\begin{eqnarray}
 \Sigma_{L(R)}(F) &=& \int_{C_{\rm left(right)}}\frac{dz}{2\pi i}
F(z)\,\sigma(z).
\end{eqnarray}
Similarly, we can obtain a generic formula associated with the identity
string field $I$:
\begin{eqnarray}
\label{Eq:LRI}
 \Sigma_L(F)I=-\Sigma_R(F)I.
\end{eqnarray}
If we choose $c(z)$, $v(z)$ and $\tilde{v}(z)$ as the primary
field,\footnote{The dimensions of $c(z)$, $v(z)$ and $\tilde{v}(z)$ are
$-1$, $0$ and $0$, respectively.} we can derive the following equations
from the generic formulae:
\begin{eqnarray}
\label{Eq:Cstar}
  \left(C_R(F_{-1}) A\right)*B&=&-(-1)^{\abs{A}}
A*\left(C_L(F_{-1})B\right),\\
\label{Eq:Vstar}
  \left(V_R(F_0) A\right)*B&=&-(-1)^{\abs{A}}
A*\left(V_L(F_0)B\right),\\
\label{Eq:Vtildstar}
  \left(\tilde{V}_R(F_0) A\right)*B&=&-
A*\left(\tilde{V}_L(F_0)B\right),\\
\label{Eq:CI}
  C_R(F_{-1}) I&=&- C_L(F_{-1}) I,\\
\label{Eq:VI}
  V_R(F_0) I&=&- V_L(F_0) I,\\
\label{Eq:VtildI}
  \tilde{V}_R(F_0) I&=&- \tilde{V}_L(F_0) I,
\end{eqnarray}
where the function $F_h(z)$ satisfies $F_h(-1/z)=(z^2)^{1-h}F_h(z)$.
Again, these equations hold if 
$F_h(z)$ is defined on the unit circle $\abs{z}=1$. As in the case of
eqs.~(\ref{Eq:VLVtildL}) and (\ref{Eq:VLVL}), the function
does not need holomorphicity. 
Here, the function
in eq.~(\ref{Eq:CI}) should behave like $F_{-1}(z)\sim {\rm
O}((z-z_0)^\epsilon),\ (\epsilon>0)$
near $z_0=\pm i$ in order that the state 
$C_{L(R)}(F_{-1})I$ has a well-defined Fock space expression.
Because the ghost field $c(z)$ has a single pole at $z=\pm i$ on the
identity state as seen in the next subsection
\cite{rf:TT1,rf:TT2,rf:tomo}. This condition is assured
by imposing $F_{-1}(\pm i)=0$ if the function is expandable in a Taylor
series.\footnote{Actually, we can expand the function as
$F_{-1}(z)=O((z-z_0)^2)$ if $F_{-1}(\pm i)=0$ holds.}

Now, it can be easily shown that $\Phi_0$ given by (\ref{Eq:sol}) is a
classical solution:
\begin{eqnarray}
 e^{-\Phi_0}*\Q\,e^{\Phi_0} &=&
e^{\tilde{V}_L(F)}\,\Q\,e^{-\tilde{V}_L(F)}\,I\\
&=& \Q I+
\left[\tilde{V}_L(F),\,\Q\right] I+\frac{1}{2!}
\left[\tilde{V}_L(F),\,\left[\tilde{V}_L(F),\,\Q\right]\right]I+\cdots\\
\label{Eq:PhiQPhi}
&=& -V_L(F)I+\frac{1}{4}\,C_L(F^2)I,
\end{eqnarray}
where we have used (\ref{Eq:VtildI}) and
$\left[\tilde{V}_L(F),\,\tilde{V}_R(F)\right]=0$ in the first equality,
and eq.~(\ref{Eq:VLVtildL}) in the last equality.
We should note that the state $C_L(F^2)I$ is well-defined
because $F^2(\pm i)=0$ due to $F(-1/z)=z^2 F(z)$.
The $\xi$ zero mode is not contained in both operators $V_L(F)$ and
$C_L(F^2)$ and the identity string field satisfies $\eta_0 I=0$.
As a result, we find that $\eta_0(e^{-\Phi_0}*\Q e^{\Phi_0})=0$ and then
$\Phi_0$ is a solution in open superstring field theory.

%%%%%%%%%%%%%%%%%%%%%%%%%%%%%%%%%%%%%%%%%%%%%%%%%%%%%%%%%%%%%%%%
\subsection{Fock space expressions}

The operator $\tilde{V}_L(F)$ in the solution can be expressed
in terms of integration with respect to $\sigma$
($z=e^{i\sigma}$ on the contour $C_{\rm left}$) :
\begin{eqnarray}
 \tilde{V}_L(F)&=&\int_{-{\pi\over 2}}^{\pi\over 2} {d\sigma\over 2\pi}
{e^{i\sigma}F(e^{i\sigma})\over \sqrt{2}}c(i\sigma)
\xi(i\sigma) :\!e^{-\phi(i\sigma)}\!:\psi(i\sigma)\,,
\label{eq:VLFOsc}
\end{eqnarray}
where operators in the integrand are given by oscillator
expansions:
\begin{eqnarray}
&&c(i\sigma)=\sum_{n\in \mathbf{Z}}c_ne^{-in\sigma},
~~\xi(i\sigma)=\sum_{n\in \mathbf{Z}}\xi_ne^{-in\sigma},~~
\psi(i\sigma)=\sum_{r\in \mathbf{Z}+{1\over 2}}\psi_re^{-ir\sigma}\,,
\end{eqnarray}
and $:\!e^{-\phi(i\sigma)}\!:$ is expanded in eq.~(\ref{eq:eqphi}). Using
formulae: eqs.~(\ref{eq:conn2}), (\ref{eq:conn3}) and
(\ref{eq:conn4}) and computing straightforwardly, they are expressed 
on the identity state $|I\rangle$ as
\begin{eqnarray}
 c(i\sigma)|I\rangle&=&
\biggl[c_1(2\cos\sigma)^{-1}+c_0i\tan\sigma
+c_{-1}\left(1+(2\cos\sigma)^{-1}\right)\nonumber\\
&&~~~+2\sum_{k\ge 1}\left(c_{-2k}i\sin
		     2k\sigma+ c_{-(2k+1)}\cos(2k+1)\sigma
\right)\biggr]|I\rangle\,,\\
\xi(i\sigma)|I\rangle&=&\left[\xi_0+2\sum_{k\ge 1}
\left(\xi_{-2k}\cos 2k\sigma +\xi_{-(2k-1)}i\sin(2k-1)\sigma\right)
\right]|I\rangle\,,\\
:\!e^{-\phi(i\sigma)}\!:|I\rangle&=&(2\cos\sigma)^{1\over 2}\,
e^{\sum_{k\ge 1}\left({\cos 2k\sigma\over
		 k}j_{-2k}+{2i\sin(2k-1)\sigma\over
		 2k-1}j_{-(2k-1)}\right)}
e^{-\hat{\phi}_0}|I\rangle\,,
\label{eq:ephiI}\\
\psi(i\sigma)|I\rangle&=&\sum_{r,s\ge 1/2}\left(
\delta_{r,s}e^{is\sigma}+e^{-ir\sigma}I_{rs}
\right)\psi_{-s}|I\rangle\nonumber\\
&=&\sum_{n=0}^{\infty}(2\cos\sigma)^{1\over 2}\oint_0{dz\over 2\pi i}
z^{-n-1}{\sqrt{1+z^2}\over (1-e^{i\sigma}z)(1+e^{-i\sigma}z)}
\psi_{-(n+{1\over 2})}|I\rangle\nonumber\\
&=&{\sqrt{2}\over (\cos\sigma)^{1\over
 2}}\sum_{k=0}^{\infty}\biggl[
\psi_{-(2k+{1\over 2})}\sum_{m=0}^k{(-1)^{k-m+1}(2(k-m))!\over 
2^{2(k-m)}((k-m)!)^2(2(k-m)-1)}\cos(2m+1)\sigma\nonumber\\
&&+\psi_{-(2k+{3\over 2})}\sum_{m=0}^k{(-1)^{k-m+1}(2(k-m))!\over 
2^{2(k-m)}((k-m)!)^2(2(k-m)-1)}i\sin 2(m+1)\sigma
\biggr]|I\rangle\,.
\label{eq:psiI}
\end{eqnarray}
In computing $\psi(i\sigma)|I\rangle$, we have used 
a relation for the Neumann coefficients:
\begin{eqnarray}
&&\sum_{r,s\ge {1\over 2}}I_{rs}y^{r-{1\over 2}}z^{s-{1\over 2}}
={(h'_I(y))^{1\over 2}
(h'_I(z))^{1\over 2}\over h_I(y)-h_I(z)}-{1\over y-z}
={\sqrt{1+y^2}\sqrt{1+z^2}\over (y-z)(1+yz)}-{1\over y-z}\,,
\end{eqnarray}
where $h_I(z)=2z/(1-z^2)$ is the conformal map for
the identity string field (see appendix \ref{sec:Osc}).
Using the above expressions and the reality ${\rm
bpz}(|I\rangle)=(|I\rangle)^{\dagger}$, we find a relation between the
BPZ and hermitian conjugations:
\begin{eqnarray}
 {\rm
bpz}(c(i\sigma)\xi(i\sigma):\!e^{-\phi(i\sigma)}\!:\psi(i\sigma)|I\rangle)
=-(c(i\sigma)\xi(i\sigma):\!e^{-\phi(i\sigma)}\!:\psi(i\sigma)
|I\rangle)^{\dagger},
\end{eqnarray}
where we take a convention:
$c^{\dagger}_n=c_{-n},\xi^{\dagger}_{n}=-\xi_{-n},
j_n^{\dagger}=-j_{-n}-2\delta_{n,0},\psi_r^{\dagger}=\psi_{-r}$.
As a result, the reality condition for our solution ${\rm
bpz}(|\Phi_0\rangle)=(|\Phi_0\rangle)^{\dagger}$ imposes 
$(e^{i\sigma}F(e^{i\sigma}))^*=-e^{i\sigma}F(e^{i\sigma})$
for the coefficient function in the integrand of (\ref{eq:VLFOsc}),
which is expanded as
\begin{eqnarray}
 e^{i\sigma}F(e^{i\sigma})&=&\sum_{n\in {\mathbf{Z}}}F_ne^{-in\sigma}
=\sum_{n\ge 1}F_{-n}(e^{in\sigma}-(-1)^ne^{-in\sigma})\nonumber\\
&=&2\sum_{k=1}^{\infty}\left(F_{-2k}i\sin 2k\sigma
+F_{-(2k-1)}\cos(2k-1)\sigma\right).
\end{eqnarray}
Here we have used $F_n=-(-1)^nF_{-n}$ in the second equality,
which follows from the condition $F(-1/z)=z^2F(z)$ for 
the classical solution.
The reality condition implies that $F_{-2k}$ should be real and 
$F_{-(2k-1)}$ should be pure imaginary.

Putting the above expansions together,
 we obtain the explicit Fock space expression of the classical solution
as follows:
\begin{eqnarray}
|\Phi_0\rangle&=&-\tilde{V}_L(F)|I\rangle\nonumber\\
&=&\sqrt{2}\sum_{n=1}^{\infty}
\int_{-{\pi\over 2}}^{\pi\over 2} {d\sigma\over \pi}
\left(F_{-2n}i\sin 2n\sigma+F_{-(2n-1)}\cos(2n-1)\sigma\right)
\nonumber\\
&&~~~~\times\biggl[c_1(2\cos\sigma)^{-1}+
c_0i\tan\sigma
+c_{-1}\left(1+(2\cos\sigma)^{-1}\right)\nonumber\\
&&~~~~~~~~~~~~+2\sum_{m\ge 1}\left(c_{-2m}i\sin
		     2m\sigma+ c_{-(2m+1)}\cos(2m+1)\sigma
\right)\biggr]\nonumber\\
&&~~~~\times\left[\xi_0+2\sum_{l\ge 1}
\left(\xi_{-2l}\cos 2l\sigma +\xi_{-(2l-1)}i\sin(2l-1)\sigma\right)
\right]\\
&&~~~~\times \exp\!\left[\sum_{p\ge 1}\biggl({\cos 2p\sigma\over
		 p}j_{-2p}+{2i\sin(2p-1)\sigma\over
		 2p-1}j_{-(2p-1)}\biggr)\right] e^{-\hat{\phi}_0}
\nonumber\\
&&~~~~\times\sum_{k=0}^{\infty}\biggl[
\psi_{-(2k+{1\over 2})}\sum_{q=0}^k{(-1)^{k-q}(2(k-q))!\over 
2^{2(k-q)}((k-q)!)^2(2(k-q)-1)}\cos(2q+1)\sigma\nonumber\\
&&~~~~~~~~~~~~~~
+\psi_{-(2k+{3\over 2})}\sum_{q=0}^k{(-1)^{k-q}(2(k-q))!\over 
2^{2(k-q)}((k-q)!)^2(2(k-q)-1)}i\sin2(q+1)\sigma
\biggr]|I\rangle\,.~~~~~\nonumber
\end{eqnarray}
In particular, the integration with respect to $\sigma$ gives finite
coefficients for each term of the form
$F_{-n}c_{-m}\xi_{-l}j_{-p_1}^{n_1}\cdots
j_{-p_N}^{n_N}\psi_{-s}|I\rangle$
because both $\int_{-{\pi\over 2}}^{\pi\over 2} {d\sigma\over
2\pi}\left|{\sin 2k\sigma\over \cos\sigma}\right|$ and  
$\int_{-{\pi\over 2}}^{\pi\over 2} {d\sigma\over 2\pi}
\left|{\cos(2k-1)\sigma\over \cos\sigma}\right|$,\footnote{
Notice that coefficients of {\it each} term can be estimated as
$M\left|\int_{-{\pi\over 2}}^{\pi\over 2} {d\sigma\over 2\pi}
{\sin 2k\sigma\over \cos\sigma}\sin^{m_1}\!n_1\sigma\cdots
\cos^{p_1}\!q_1\sigma\cdots \right|<
M\int_{-{\pi\over 2}}^{\pi\over 2} {d\sigma\over
2\pi}\left|{\sin 2k\sigma\over \cos\sigma}\right|$
or 
$M'\left|\int_{-{\pi\over 2}}^{\pi\over 2} {d\sigma\over 2\pi}
{\cos (2k-1)\sigma\over \cos\sigma}\sin^{m_1'}\!n_1'\sigma\cdots
\cos^{p_1'}\!q_1'\sigma\cdots \right|<
M'\int_{-{\pi\over 2}}^{\pi\over 2} {d\sigma\over
2\pi}\left|{\cos (2k-1)\sigma\over \cos\sigma}\right|$,
where $M,M'$ are some finite positive constants.
}
in which the numerators and the denominators come from
the coefficients $e^{i\sigma}F(e^{i\sigma})$ and 
zero mode of ghost $c$ (i.e., $c_1,c_0,c_{-1}$) respectively, are finite.
We note that a factor $(\cos \sigma)^{-{1\over 2}}$ in (\ref{eq:psiI})
is canceled by a factor $(\cos\sigma)^{1\over 2}$ in (\ref{eq:ephiI}).
Therefore, one can construct well-behaved solutions
in the sense that coefficients of all  modes in the Fock space
become finite by taking appropriate $F(z)$. A sufficient condition 
is that only finite number of $F_{-n}$s have nonzero value.

More concretely, the lowest few terms of the solution are computed as
\begin{eqnarray}
 |\Phi_0\rangle&=&-{\sqrt{2}\over \pi}
\biggl(\sum_{n=1}^{\infty}
{(-1)^nF_{-(2n-1)}\over 2n-1}c_1\xi_0\psi_{-{1\over 2}}\\
&&
+\sum_{n=1}^{\infty}
{(-1)^n4 nF_{-2n}\over  4n^2-1}\left(
(c_0\xi_0+c_1\xi_{-1}+c_1\xi_0j_{-1})\psi_{-{1\over 2}}
+c_1\xi_0\psi_{-{3\over 2}}\right)+
 \cdots \biggr)e^{-\hat{\phi}_0}|I\rangle.\nonumber
\end{eqnarray}
In the above explicit expression, 
the first term implies the condensation of the massless
vector field because it is expanded as
$c_1\xi_0\psi e^{-\hat{\phi}_0}|I\rangle\sim 
-c\xi e^{-\phi}\psi(0)|0\rangle+\cdots $ and $c\xi e^{-\phi}\psi$
is the vertex operator for massless vector with zero
momentum \cite{rf:BS}.
This coefficient constant for the lowest level can be rewritten as
$-{\sqrt{2}\over \pi}\sum_{n=1}^{\infty}
{(-1)^nF_{-(2n-1)}\over 2n-1}=\int_{C_{\rm left}}{dz\over
2\pi i}{F(z)\over \sqrt{2}}$.

%%%%%%%%%%%%%%%%%%%%%%%%%%%%%%%%%%%%%%%%%%%%%%%%%%%%%%%%%%%%%%%%
\subsection{background Wilson lines}

We found that the classical solution involves the condensation of the
massless vector field. This result suggests that our solution is related
to a background Wilson line.
In this subsection, we will discuss the vacuum energy of the classical
solution, the theory expanded around the solution, and other
characteristic features 
of the solution. Accordingly, we will show that the solution corresponds
to a background Wilson line.

In order to evaluate the vacuum energy, it is convenient to use an
alternative expression for the action:
\begin{eqnarray}
\label{Eq:action2}
 S[\Phi]&=&-\frac{1}{g^2}\int_0^1 dt
\Bra (\eta_0\,e^{-t\Phi} \partial_t e^{t\Phi})
(e^{-t\Phi}\,\Q e^{t\Phi}) \Ket. \\
&=&-\frac{1}{g^2}\int_0^1 dt
\Bra (\eta_0\,\Phi)
(e^{-t\Phi}\,\Q e^{t\Phi}) \Ket.
\end{eqnarray}
The equivalence of the actions (\ref{Eq:action}) and (\ref{Eq:action2})
is proved in ref.~\cite{rf:BOZ}. 
In general, the state $\eta_0\Phi$ has no $\xi$ zero mode. 
For the solution (\ref{Eq:sol}), it is easily seen that
\begin{eqnarray}
e^{-t \Phi_0}*\Q e^{t \Phi_0}=-tV_L(F)I+\frac{t^2}{4}C_L(F^2)I,
\end{eqnarray}
and then the state $e^{-t \Phi_0}*\Q e^{t \Phi_0}$ also does not
contain the $\xi$ zero mode.
As a result, we find that the integrand in (\ref{Eq:action2}) becomes
zero 
for the classical solution since there is no $\xi$ zero mode in the
correlation function of 
the integrand.\footnote{In the large Hilbert space, correlation
functions are normalized as to be 
$\left\langle\!\left\langle c\partial c \partial^2 c \xi
e^{-2\phi} \right\rangle\!\right\rangle \neq 0$.
}
Hence we confirm that the vacuum energy of the solution vanishes due
to the ghost charge non-conservation in the large Hilbert space.

Let us consider the expansion of the string field around the solution
(\ref{Eq:sol}). 
Generally, if we expand the string field $\Phi$ around a classical
solution $\Phi_0$ as $e^\Phi=e^{\Phi_0}\,e^{\Phi'}$,
the action becomes
\begin{eqnarray}
 S[\Phi] &=&  S[\Phi_0]+S'[\Phi'],\\
\label{Eq:expdAction}
 S'[\Phi'] &=& 
\frac{1}{2g^2} \Bra\,
(e^{-\Phi'} \Q' e^{\Phi'})(e^{-\Phi'}\eta_0 e^{\Phi'})
\nn
&&\ \ -\int_0^1 dt\,(e^{-t\Phi'}\partial_t e^{t\Phi'})
\left\{(e^{-t\Phi'}\Q' e^{t\Phi'}),\,
(e^{-t\Phi'}\eta_0 e^{t\Phi'})\right\}
\Ket,
\end{eqnarray}
where $S[\Phi_0]$ corresponds to the vacuum energy and 
$S'[\Phi']$ has the same form as the original
action (\ref{Eq:action}) except the kinetic operator $\Q'$, which is
defined as
\begin{eqnarray}
\label{Eq:newQB}
 \Q' \Psi=\Q\Psi+A_0*\Psi-(-1)^{\abs{\Psi}}\Psi*A_0,\ \ \ 
 A_0=e^{-\Phi_0}*\Q e^{\Phi_0}\ \ \ {\rm for}\ {}^\forall \Psi.
\end{eqnarray}
A proof is given in appendix \ref{sec:aacs}. 
For the new BRS charge, nilpotency
holds automatically but $\{\Q',\,\eta_0\}=0$ is satisfied owing to the
equation of motion, $\eta_0\,A_0=0$.
For the solution (\ref{Eq:sol}), we find $S[\Phi_0]=0$ as evaluated
above. Substituting 
(\ref{Eq:sol}) into (\ref{Eq:newQB}) and using 
(\ref{Eq:Cstar}), (\ref{Eq:Vstar}), (\ref{Eq:CI}), (\ref{Eq:VI}) and
(\ref{Eq:PhiQPhi}), we can write the new BRS charge
as\footnote{We can check nilpotency of the new BRS charge 
in terms of (\ref{Eq:VLVL}).} 
\begin{eqnarray}
 \Q'=\Q-(V_L(F)+V_R(F))+\frac{1}{4}(C_L(F^2)+C_R(F^2)).
\end{eqnarray}
Using (\ref{Eq:VLVtildL}), the new BRS charge is rewritten as a
similarity transformation from the original BRS charge:
\begin{eqnarray}
 \Q'=e^{\tilde{V}_L(F)+\tilde{V}_R(F)}\,\Q\,
e^{-\tilde{V}_L(F)-\tilde{V}_R(F)}.
\end{eqnarray}

Here, we introduce the following half string operators,
\begin{eqnarray}
 X_{L(R)}(F) &=& \int_{C_{\rm left(right)}}\frac{dz}{2\pi i}
F(z) X(z),\\
\Omega_{L(R)}(F) &=& 
\int_{C_{\rm left(right)}}\frac{dz}{2\pi i}
F(z) \,i\,c\xi \partial \xi e^{-2\phi}\,X(z).
\end{eqnarray}
Using a similar procedure in the previous subsection,
we can obtain (anti-)commutation relations between these operators in
terms of their OPEs:
\begin{eqnarray}
\label{Eq:XLVL}
 [X_{L(R)}(F),\,V_{L(R)}(F)]&=&i\sqrt{\alpha'}C_{L(R)}(F^2),\\
\label{Eq:QBXL}
 \left[\Q,\,X_{L(R)}(F)\right]&=&-i2\sqrt{\alpha'} V_{L(R)}(F),\\
\label{Eq:QBOmegaL}
 \{\Q,\,\Omega_{L(R)}(F)\}&=&
   2\sqrt{\alpha'}\tilde{V}_{L(R)}(F)-iX_{L(R)}(F).
\end{eqnarray}
If $F(z)$ satisfies $F(-1/z)=z^2 F(z)$, it follows from
(\ref{Eq:LRstar}) and (\ref{Eq:LRI}) that
\begin{eqnarray}
\label{Eq:XLRstar}
&& (X_R(F)A)*B=-A*(X_L(F)B),\ \ \ X_R(F)I=-X_L(F)I,\\
\label{Eq:OmegaLRstar}
&& (\Omega_R(F)A)*B=-(-1)^{\abs{A}}A*(\Omega_L(F)B),
\ \ \ \Omega_R(F)I=-\Omega_L(F)I.
\end{eqnarray}
Precisely speaking,
$X(z)$ is not a primary field and we can not apply the formula
(\ref{Eq:LRstar}) for the case that $\sigma(z)=X(z)$.  
However, it is directly shown that the equation (\ref{Eq:psiAB}) holds
for $X(z)$ \cite{rf:GJ1,rf:IOS, Ohta} and then we can derive the same formula
in which $X(z)$ behaves like a primary field with dimension 0 on the
string vertex. The same
holds for the formula associated with the identity string field.

In the theory expanded around the solution (\ref{Eq:sol}), we redefine
the string field $\Phi'$ as
\begin{eqnarray}
\label{Eq:redef}
 \Phi'
&=& \exp\left(\frac{i}{2\sqrt{\alpha'}}X_L(F)I\right)\,*\Phi''*\,
        \exp\left(-\frac{i}{2\sqrt{\alpha'}}X_L(F)I\right)\nn
 &=& \exp\left(\frac{i}{2\sqrt{\alpha'}}
(X_L(F)+X_R(F))\right)\,\Phi''.
\end{eqnarray}
Under this redefinition, the action of $\Phi'$ is transformed to the
exactly same form as the original action, because
$\Q'$ is transformed to $\Q$: 
\begin{eqnarray}
 \Q&=&\exp\left(-\frac{i}{2\sqrt{\alpha'}}(X_L(F)+X_R(F))
\right)\,\Q'\,
\exp\left(\frac{i}{2\sqrt{\alpha'}}(X_L(F)+X_R(F))\right),
\end{eqnarray}
where use has been made of (\ref{Eq:XLVL}) and (\ref{Eq:QBXL}).
This equivalence between the original and expanded actions suggests that
$\Phi_0$ is a pure gauge solution. Actually, 
we can represent the solution as a pure gauge form by using
(\ref{Eq:QBOmegaL}), (\ref{Eq:XLRstar}) and (\ref{Eq:OmegaLRstar}):
\begin{eqnarray}
\label{eq:local_pure}
 e^{\Phi_0}=
\exp\left\{
\Q\left(-\frac{1}{2\sqrt{\alpha'}}\,\Omega_L(F)I\right)
\right\}*
\exp\left\{\eta_0
\left(-\frac{i}{2\sqrt{\alpha'}}\xi_0X_L(F)I\right)\right\}.
\end{eqnarray}
However, this is merely a locally pure gauge expression,
because the operator $X_L(F)$ contains the zero mode $\hat{x}$ which can not
be defined globally 
if the direction is compactified. Hence, the classical 
solution turns out to be non-trivial.

In order to clarify the physical meaning of the solution, let us
consider the case involving the Chan-Paton factor represented
with indices $(i,\,j)$. 
The string field redefinition (\ref{Eq:redef}) can
be generalized to\footnote{The function $F_i(z)$ corresponds to
$\lambda_i F(z)$ in ref.~\cite{rf:KTZ}.} 
\begin{eqnarray}
\label{Eq:redefij}
 \Phi'_{ij}
&=& \exp\left(\frac{i}{2\sqrt{\alpha'}}X_L(F_i)I\right)\,*
\Phi''_{ij}*\,
\exp\left(-\frac{i}{2\sqrt{\alpha'}}X_L(F_j)I\right)\nn
 &=& \exp\left(\frac{i}{2\sqrt{\alpha'}}
X_L(F_i)+\frac{i}{2\sqrt{\alpha'}}
X_R(F_j)\right)\,\Phi''_{ij},
\end{eqnarray}
where we take no sum with respect to $(i,\,j)$. Noting
\begin{eqnarray}
 \int_{C_{\rm left}}\frac{dz}{2\pi i}F(z)=
 -\int_{C_{\rm right}}\frac{dz}{2\pi i}F(z)\ \ \ 
{\rm for}\ \ F(-1/z)=z^2 F(z),
\end{eqnarray}
it is expanded as
\begin{eqnarray}
 \Phi''_{ij}=
\exp\left(-\frac{i}{2\sqrt{\alpha'}}(f_i-f_j)\hat{x}
+\cdots\right)\Phi'_{ij},
\ \ \ f_i= \int_{C_{\rm left}}\frac{dz}{2\pi i}F_i(z),
\end{eqnarray}
where the abbreviation on the exponent denotes nonzero mode dependence.
Consequently, this string field redefinition causes shift of momentum
$p\rightarrow p-(f_i-f_j)/(2\sqrt{\alpha'})$ which is the
same effect by background Wilson lines as shown in bosonic string field
theory \cite{rf:TT1}. 

Thus, we conclude that the classical solution (\ref{Eq:sol}) corresponds
to background Wilson lines, because the vacuum energy vanishes and
the solution is represented as locally pure gauge form, and the theory
expanded around the solution involves the momentum shift as expected from
Wilson lines.

%%%%%%%%%%%%%%%%%%%%%%%%%%%%%%%%%%%%%%%%%%%%%%%%%%%%%%%%%%%%%%%%
\subsection{Ramond sector and supersymmetry}

The action of the Ramond sector proposed in ref.~\cite{rf:Michi} is:
\begin{eqnarray}
\label{Eq:actionF}
 S_F=
  -\frac{1}{2g^2}\Bra (\Q\Xi) e^\Phi (\eta_0\Psi)e^{-\Phi}\Ket,
\end{eqnarray}
where $\Psi$ is a string field of the GSO(+) R sector which has ghost
number 0 and picture number $1/2$. $\Xi$ carries ghost number 0 and
picture number $-1/2$. Both $\Psi$ and $\Xi$ are Grassmann even.
The total action is given by adding $S_F$ to the NS action
(\ref{Eq:action}), and a constraint is imposed on string fields as
\begin{eqnarray}
\label{Eq:constraint}
 \Q\Xi = e^\Phi(\eta_0\Psi)e^{-\Phi}.
\end{eqnarray}
The total action is invariant under the infinitesimal gauge
transformation 
\begin{eqnarray}
\label{Eq:infgaugetrans}
 \delta e^{\Phi}&=&(\Q\delta\Lambda_0)*e^\Phi+
e^\Phi*(\eta_0\delta\Lambda_1),\\
\label{eq:infgaugetrans2}
 \delta \Psi &=& \eta_0\delta\Lambda_{3/2}
+\Psi*(\eta_0\delta\Lambda_1)-(\eta_0\delta\Lambda_1)*\Psi,\\
\label{eq:infgaugetrans3}
 \delta \Xi &=&\Q\delta\Lambda_{-1/2}
+(\Q\delta\Lambda_0)*\Xi-\Xi*(\Q\delta\Lambda_0),
\end{eqnarray}
where $\delta\Lambda_P$ denotes an infinitesimal parameter with the picture
number $P$.
The constraint (\ref{Eq:constraint}) is unchanged under the
transformation.

Variating the total action, we can derive the
equations of motion to be \cite{rf:Michi}
\begin{eqnarray}
\label{Eq:eomNS}
 \eta_0(e^{-\Phi}\,\Q e^\Phi)&=& 
  -\frac{1}{2}(\eta_0\Psi)e^{-\Phi}(\Q\Xi)e^\Phi
  -\frac{1}{2}e^{-\Phi}(\Q\Xi)e^\Phi(\eta_0\Psi),\\
\label{Eq:eomXi}
 \eta_0(e^{-\Phi}(\Q\Xi)e^\Phi)&=& 0,\\
\label{Eq:eomR}
 \Q(e^\Phi(\eta_0\Psi)e^{-\Phi})&=& 0.
\end{eqnarray}
Substituting the constraint (\ref{Eq:constraint}) into these equations,
we can obtain the equations of motion for the NS and R sectors, 
$ \eta_0(e^{-\Phi}\,\Q e^\Phi)=-(\eta_0\Psi)^2 $ and
$\Q(e^\Phi(\eta_0\Psi)e^{-\Phi})= 0$ \cite{rf:Rsector}. 

Let us expand the string fields around a classical solution
$(\Phi,\,\Psi)=(\Phi_0,\,0)$ as
($e^\Phi,\,\Psi)=(e^{\Phi_0}e^{\Phi'},\,\Psi')$. Then, the 
action of the R sector becomes
\begin{eqnarray}
\label{Eq:actionFexp}
 S_F=
  -\frac{1}{2g^2}\Bra (\Q'\Xi') e^{\Phi'} (\eta_0\Psi')e^{-\Phi'}\Ket,
\end{eqnarray}
and the constraint is changed to
\begin{eqnarray}
 \Q'\Xi'=e^{\Phi'}(\eta_0\Psi')e^{-\Phi'},
\end{eqnarray}
where $\Q'$ is the new BRS operator defined as (\ref{Eq:newQB}) and
$\Xi$ is a superfluous string field redefined as
$\Xi'=e^{-\Phi_0}\Xi e^{\Phi_0}$. Like the NS sector, the
expanded action and constraint in the R sector have the same structure
as the original ones except the form of the BRS operator.

Now, we take the solution (\ref{Eq:sol}) as $\Phi_0$ in
the above expansion. In the expanded action,  we redefine the string
fields as 
\begin{eqnarray}
\label{Eq:redefsuper}
 \Phi'
&=& \exp\left(\frac{i}{2\sqrt{\alpha'}}X_L(F)I\right)\,*\Phi''*\,
        \exp\left(-\frac{i}{2\sqrt{\alpha'}}X_L(F)I\right),\\
\Psi' &=& 
 \exp\left(\frac{i}{2\sqrt{\alpha'}}X_L(F)I\right)\,*\Psi''*\,
        \exp\left(-\frac{i}{2\sqrt{\alpha'}}X_L(F)I\right),\\
\Xi' &=& 
 \exp\left(\frac{i}{2\sqrt{\alpha'}}X_L(F)I\right)\,*\Xi''*\,
        \exp\left(-\frac{i}{2\sqrt{\alpha'}}X_L(F)I\right).
\end{eqnarray}
We can easily find that the new BRS charge
is transformed to the original form in both of the total action and the
constraint. As in the previous subsection, the classical
solution has an effect only on the momentum in the string
fields. The result indicates that the classical solution (\ref{Eq:sol})
corresponds to the background Wilson line in open superstring field
theory including the NS and R sector.

Instead of using the action, we can see the effect of the classical
solution in terms of the equation of motion.
The original Berkovits' equations of motion are given by
$\eta_0(e^{-\Phi}\Q e^{\Phi})=-(\eta_0\Psi)^2,~
\Q(e^{\Phi}(\eta_0\Psi)e^{-\Phi})=0$. Expanding the equations of motion
around the classical solution,
we find that the form of the equations is unchanged but the BRS charge
is changed to $\Q'$ of (\ref{Eq:newQB}).
In the case of our solution $\Phi_0=-{\tilde V}_L(F)I,\Psi_0=0$
(\ref{Eq:sol}), the expanded equations of motion
are transformed back to the original ones by the field redefinitions:
$\Phi'=e^{{i\over 2\sqrt{\alpha'}}X_L(F)I}*\Phi''*e^{-{i\over
2\sqrt{\alpha'}}X_L(F)I},~
\Psi'=e^{{i\over 2\sqrt{\alpha'}}X_L(F)I}*\Psi''*e^{-{i\over
2\sqrt{\alpha'}}X_L(F)I}$. This redefinition reflects the effect of the
Wilson lines. 

The equations of motion for the NS and R sectors have a fermionic gauge
symmetry as follows \cite{rf:Rsector}:
\begin{eqnarray}
\label{Eq:fermionicsym}
 \delta e^\Phi &=& -e^{\Phi}*(\eta_0\Psi*\Lambda_{1/2}
+\Lambda_{1/2}*\eta_0\Psi),\\
\delta \Psi &=& \Q\Lambda_{1/2}+e^{-\Phi}\Q e^\Phi*\Lambda_{1/2}
+\Lambda_{1/2}*e^{-\Phi}\Q e^\Phi,
\label{eq:fermionic_g2}
\end{eqnarray}
where $\Lambda_{1/2}$ is a Grassmann odd parameter 
with the picture number $1/2$. 
This may include a global space-time supersymmetry, since the NS (R)
string field is transformed to the R (NS) sector under the
transformation. Actually, we can find the supersymmetry if we 
formally set a transformation parameter as 
\begin{eqnarray}
\label{eq:Oep}
 \Lambda_{1/2}=\Omega(\epsilon)=
 \epsilon_\alpha \int_{C_{\rm left}}{dz\over 2\pi i}\xi
 S^\alpha_{(-1/2)}(z) I,
\end{eqnarray}
where $S^\alpha_{(-1/2)}$ is a GSO(+) spin field with $\phi$-charge
$-1/2$ and positive chirality
and $\epsilon_\alpha$ is a fermionic constant.
We give the details of the supersymmetry in appendix~C.
Substituting (\ref{eq:Oep}) into eqs.~(\ref{Eq:fermionicsym}) and
(\ref{eq:fermionic_g2}),
we can rewrite the transformation law as\footnote{
\label{footnote:susy}
We do not include a contribution from the first term in
(\ref{eq:fermionic_g2}): $\delta_{\rm M}(\eta_0\Psi)\equiv
-\Q\eta_0\Omega(\epsilon)={i\over 2\pi}
\epsilon_{\alpha}(cS^{\alpha}_{(-1/2)}(i)-cS^{\alpha}_{(-1/2)}(-i))I$.
The elimination of this term is possible
because the transformation 
$\delta_{\rm M}$ is a symmetry of the equations of motion:
$\delta_{\rm M}f_1=\{{\cal O}(i)I,\eta_0\Psi\}={\cal O}(i)
\eta_0\Psi-{\cal O}(i)
\eta_0\Psi=0,~\delta_{\rm M}f_2=\{{\cal O}(i)I,e^{-\Phi}\Q e^{\Phi}\}
=0$, where $f_1$ and $f_2$ are given in eqs.~(\ref{eq:EOMf1}) and
(\ref{eq:EOMf2}) respectively and  ${\cal O}(i)\equiv{i\over 2\pi}
\epsilon_{\alpha}(cS^{\alpha}_{(-1/2)}(i)-cS^{\alpha}_{(-1/2)}(-i))$.
}
\begin{eqnarray}
\label{Eq:susytrans}
 \delta_\epsilon e^\Phi=-e^\Phi {\cal S}(\epsilon)\eta_0 \Psi,
\ \ \ \delta_\epsilon (\eta_0\Psi) =
\eta_0{\cal S}(\epsilon)(e^{-\Phi}\Q e^\Phi),
\end{eqnarray}
where the operator ${\cal S}(\epsilon)$ is defined as
\begin{eqnarray}
\label{eq:Sepsilon}
 {\cal S}(\epsilon)=\epsilon_\alpha
 \oint\frac{dz}{2\pi i} \xi S^\alpha_{(-1/2)}(z).
\end{eqnarray}
The operator ${\cal S}(\epsilon)$ is an anti-derivation with respect to the
star product.
Now, we apply this transformation to the Wilson line solution, namely
$\Phi_0$ given by (\ref{Eq:sol}) and $\Psi_0=0$:
\begin{eqnarray}
 \delta_\epsilon e^\Phi &=& 0,\\
 \delta_\epsilon (\eta_0\Psi) &=& \eta_0{\cal S}(\epsilon)(e^{-\Phi_0}
\Q e^{\Phi_0}) 
=\epsilon_\alpha  \oint\frac{dz}{2\pi i} S^\alpha_{(-1/2)}(z)
\left(V_L(F)-\frac{1}{4}C_L(F^2)\right)I \nn
&=& 
\epsilon_\alpha  \oint\frac{dz}{2\pi i} \{S^\alpha_{(-1/2)}(z),\,
V_L(F)\,\}I =0,
\end{eqnarray}
where use has been made of $\{v(y),\,S^\alpha_{(-1/2)}(z)\}=0$.
We note that $\displaystyle \oint \frac{dz}{2\pi i}
S^\alpha_{(-1/2)}I=0$ and ${\cal S}(\epsilon)I=0$ because
both $S^\alpha_{(-1/2)}(z)$ and $\xi S^\alpha_{(-1/2)}(z)$ are primary
fields with conformal dimension 1. This result indicates that the Wilson
line solution exactly preserves all global space-time supersymmetries.

%%%%%%%%%%%%%%%%%%%%%%%%%%%%%%%%%%%%%%%%%%%%%%%%%%%%%%
\subsection{symmetries and classical solutions}

The half integration of $F(z)$ is related to a Wilson line parameter,
which should be a gauge invariant quantity of the stringy gauge group.
This relation suggests that other modes of $F(z)$ are redundant degrees
of freedom under the gauge symmetry. 
In this subsection we would like to illustrate this point, namely the
function $F(z)$ can be changed by an appropriate transformation except
the half integration mode. 

The total action including the R sector is invariant under the finite
gauge transformation,
\begin{eqnarray}
\label{Eq:gaugetrans01}
 e^{\Phi^{\prime}}&=&e^{\Q\Lambda_0}*e^{\Phi}*e^{\eta_0 \Lambda_1},\\
\label{Eq:gaugetrans02}
 \Psi' &=&e^{-\eta_0 \Lambda_1}*\Psi*e^{\eta_0 \Lambda_1}
+{1-e^{-{\rm ad}_{\eta_0\Lambda_1}}\over {\rm
ad}_{\eta_0\Lambda_1}}(\eta_0\Lambda_{3/2})\,,\\
\label{Eq:gaugetrans03}
 \Xi' &=& e^{\Q \Lambda_0}*\Xi*e^{-\Q \Lambda_0}+{e^{{\rm
  ad}_{\Q\Lambda_0}}-1\over {\rm
  ad}_{\Q\Lambda_0}}(\Q\Lambda_{-1/2}).
\end{eqnarray}
This transformation can be obtained by performing $N$ times of the
infinitesimal transformation given in eqs.~(\ref{Eq:infgaugetrans}),
(\ref{eq:infgaugetrans2}) and (\ref{eq:infgaugetrans3})
and then taking
the limit $N\rightarrow \infty$:
\begin{eqnarray}
\label{Eq:gaugetrans1}
&&e^{\Phi^{\prime}}=\lim_{N\rightarrow \infty}\left(
1+{1\over N}\Q\Lambda_0\right)^N e^{\Phi}
\left(1+{1\over N}\eta_0\Lambda_1\right)^N,\\
\label{Eq:gaugetrans2}
&&\Psi'=\lim_{N\rightarrow \infty}\left[
\left(1-{1\over N}{\rm ad}_{\eta_0\Lambda_1}\right)^N\Psi
+\sum_{k=0}^{N-1}\left(1-{1\over N}{\rm ad}_{\eta_0\Lambda_1}\right)^k
{1\over N}\eta_0\Lambda_{3/2}
\right],\\
\label{Eq:gaugetrans3}
&&\Xi'=\lim_{N\rightarrow \infty}\left[
\left(1+{1\over N}{\rm ad}_{\Q\Lambda_0}\right)^N\Xi
+\sum_{k=0}^{N-1}\left(1+{1\over N}{\rm ad}_{\Q\Lambda_0}\right)^k
{1\over N}\Q\Lambda_{-1/2}
\right],
\end{eqnarray}
where we set $\delta\Lambda_P=\Lambda_P/N$ in (\ref{Eq:infgaugetrans}),
(\ref{eq:infgaugetrans2}) and (\ref{eq:infgaugetrans3})
and we have used the definition ${\rm ad}_XY\equiv [X,\,Y]=X*Y-Y*X$.

Substituting  $\Q\Lambda_0=-\eta_0 \Lambda_1=\Upsilon_0$ and 
$\Lambda_{3/2}=\Lambda_{-1/2}=0$ into (\ref{Eq:gaugetrans01}),
(\ref{Eq:gaugetrans02}) and (\ref{Eq:gaugetrans03}), 
we find 
\begin{eqnarray}
\label{Eq:simtrans}
 e^{\Phi'}=e^{\Upsilon_0}*e^{\Phi}*e^{-\Upsilon_0},~~~~
 \Psi'=e^{\Upsilon_0}*\Psi*e^{-\Upsilon_0},~~~~
 \Xi'=e^{\Upsilon_0}*\Xi*e^{-\Upsilon_0}.
\end{eqnarray}
In the large Hilbert space, any state $\Upsilon_0$ satisfying
$\Q\Upsilon_0=\eta_0\Upsilon_0=0$ can be written as 
$\Upsilon_0=\Q\Lambda_0=-\eta_0 \Lambda_1$.
Hence, the total action is invariant under the similarity
transformation generated by $\Upsilon_0$ such that
$\Q\Upsilon_0=\eta_0\Upsilon_0=0$. 

The BRS charge $\Q$ corresponds to the external derivative
in the WZW theory. It is easy to show that
the ``pure gauge connection''
$A_Q=e^{-\Phi}*\Q e^\Phi$ is transformed as
\begin{eqnarray}
 A'_Q&=&e^{-\eta_0\Lambda_1}*\Q e^{\eta_0\Lambda_1}
  +e^{-\eta_0\Lambda_1}*A_Q*e^{\eta_0\Lambda_1},
\end{eqnarray}
under the gauge transformation.
If we perform the similarity
transformation (\ref{Eq:simtrans}) on $A_Q$, we find
$A_Q'=e^{\Upsilon_0}*A_Q * e^{-\Upsilon_0}$ since $\Q e^{\Upsilon_0}=0$.
This transformation law allows us to interpret 
(\ref{Eq:simtrans}) as a ``global transformation'' \cite{rf:IIKTZ},
which is a transformation of a subgroup of the stringy gauge group.

Now, let us consider a ``global transformation'' generated by the
parameter,
\begin{eqnarray}
\label{Eq:sig0}
 \Upsilon_0={\cal T}_L(f)I=\int_{C_{\rm left}}{dz\over 2\pi i}f(z)T(z)
  I\,,~~~
f(-1/z)=z^{-2}f(z)\,,
\end{eqnarray}
where $T(z)$ is the total energy momentum tensor. It is easily seen that
$\Upsilon_0$ satisfies $\Q\Upsilon_0=\eta_0\Upsilon_0=0$ since
$[\Q,\,T(z)]=[\eta_0,\,T(z)]=0$. More explicitly, 
the gauge transformation parameters $\Lambda_0$ and $\Lambda_1$ satisfying $\Q
\Lambda_0=-\eta_0 \Lambda_1=\Upsilon_0$ can be written as
\begin{eqnarray}
&&\Lambda_0=U_L(f)I,~~~~~\Lambda_1=-\xi_0{\cal T}_L(f)I\,,
\end{eqnarray}
where $U_L(f)$ is defined as\footnote{
$T^{\rm m}(z)$ is the matter energy momentum tensor.
We note that $\{\Q,\,u(z)\}=T(z)$.}
\begin{eqnarray}
U_L(f)&=&\int_{C_{\rm left}}{dz\over 2\pi i}f(z)u(z)\,,\\
 u(z)&=&-T^{\rm m}c\xi\partial\xi e^{-2\phi}(z)
-2bc\partial c\xi\partial\xi e^{-2\phi}(z)
+c\partial\xi\partial^2\xi e^{-2\phi}(z)-{3\over 2}\partial^2 c
\xi\partial\xi e^{-2\phi}(z)\nonumber\\
&&-{1\over 2}c\xi\partial^3\xi e^{-2\phi}(z)
+c\xi\partial\xi\left(
{1\over 2}(\partial\phi)^2+3\partial^2\phi\right)e^{-2\phi}(z)\,.
\end{eqnarray}
Using some properties of ${\cal T}_L(f)$,\footnote{The operator ${\cal
T}_L(f)$ 
satisfies
\begin{eqnarray}
&&[{\cal T}_{L(R)}(f),\,{\cal T}_{L(R)}(g)]={\cal T}_{L(R)}
((\partial f)g-f\partial g),\ \ 
[{\cal T}_L(f),\,{\cal T}_R(g)]=0,\\
&&({\cal T}_R(f)A)*B=-A*({\cal T}_L(f)B),\ \ 
{\cal T}_R(f)I=-{\cal T}_L(f)I,
\end{eqnarray}
as in the bosonic case \cite{rf:IIKTZ}.}
we can rewrite the global transformation as
\begin{eqnarray}
 \Phi'=e^{{\cal T}(f)}\Phi,~~~
 \Psi'=e^{{\cal T}(f)}\Psi,~~~~
 \Xi'=e^{{\cal T}(f)}\Xi,~~~~{\cal T}(f)=\oint{dz\over 2\pi i}f(z)T(z).
\end{eqnarray}

If we apply an infinitesimal transformation on the classical
solution (\ref{Eq:sol}), it changes to
\begin{eqnarray}
 \Phi_0' &=& e^{{\cal T}(\epsilon)}\Phi_0 = -\tilde{V}_L(F')I,\\
\label{Eq:FF}
 F'(z) &=& F(z)-\partial(\epsilon(z)F(z)),
\end{eqnarray}
where we have used the commutation relation $[{\cal
T}(f),\,\tilde{V}_L(F)]= -\tilde{V}_L(\partial(f F))$.
We note that $F'(z)$ satisfies $F'(-1/z)=z^2F'(z)$. 
Accordingly, we find that
the function form of $F(z)$
are redundant under the gauge transformation.
However, we cannot change the half integration mode of $F(z)$. 
Indeed, we find that, from eq.~(\ref{Eq:FF}), 
\begin{eqnarray}
f=\int_{C_{\rm left}} \frac{dz}{2\pi i}F(z)
= \int_{C_{\rm left}} \frac{dz}{2\pi i}F'(z),
\end{eqnarray}
where we have used $\epsilon(\pm i)=F(\pm i)=0$ due to
$\epsilon(-1/z)=z^{-2}\epsilon(z)$ and $F(-1/z)=z^2 F(z)$.
These are consistent results with our expectation.
The half integration mode of $F(z)$, which is to be a physical quantity,
is invariant but other modes can be gauged away. 

%%%%%%%%%%%%%%%%%%%%%%%%%%%%%%%%%%%%%%%%%%%%%%%%%%%%%%%%%%%%%%%%
\section{Marginal deformations and classical solutions
\label{sec:super_gen}}

In the previous section, we have described a class of solutions
which correspond to the Wilson lines.
It turns out that they are based on algebra satisfied by
$u(1)$ supercurrent ${\bf{J}}(z,\theta)=\psi(z)+\theta {i\over
\sqrt{2\alpha'}}\partial X(z)$. From this
point of view, we can use the same method to construct classical
solutions of superstring field theory
 which correspond to more general supercurrents 
or marginal deformations in the context of conformal field
theory.\footnote{
As a comparison, we discuss a similar generalization in the context of 
the Witten's {\it bosonic} string field theory in 
appendix \ref{sec:bosonic}.}

Here we consider a supercurrent ${\mathbf J}^a(z,\theta)
=\psi^a(z)+\theta J^a(z)$ associated with a Lie algebra ${\cal
G}$ in the matter sector ($a=1,\cdots,{\rm dim}{\cal G}$).
In terms of component fields, we suppose that
OPE is given by
\begin{eqnarray}
\psi^a(y)\psi^b(z)&\sim&{1\over y-z}{1\over 2}\Omega^{ab}\,,
\label{eq:OPEJ1}\\
 J^a(y)\psi^b(z)&\sim&{1\over y-z}f^{ab}_{~~c}\psi^c(z)\,,
\label{eq:OPEJ2}\\
J^a(y)J^b(z)&\sim&{1\over (y-z)^2}{1\over 2}\Omega^{ab}
+{1\over y-z}f^{ab}_{~~c}J^c(z)\,,
\label{eq:OPEJ3}
\end{eqnarray}
where $f^{ab}_{~~c}$ is the structure constant of  ${\cal G}$
($f^{ab}_{~~c}=-f^{ba}_{~~c},
f^{ab}_{~~d}f^{cd}_{~~e}+f^{bc}_{~~d}f^{ad}_{~~e}+f^{ca}_{~~d}f^{bd}_{~~e}
=0$)
and $\Omega^{ab}$ is  an invertible matrix\footnote{
In the case of semi-simple Lie algebra, we can take $\Omega^{ab}$ as 
the Killing form
$\gamma^{ab}=f^{ac}_{~~d}f^{bd}_{~~c}$.
However, we have supposed the existence of invertible $\Omega^{ab}$ 
in order to 
include the cases of non-semi-simple algebra after ref.~\cite{Moh}.
} which satisfies
\begin{eqnarray}
\label{eq:Oinvariance}
 &&\Omega^{ab}=\Omega^{ba}\,,~~~~~
f^{ab}_{~~c}\Omega^{cd}+f^{ad}_{~~c}\Omega^{cb}=0\,.
\end{eqnarray}
In this case, energy momentum tensor $T(z)$ and its super partner $G(z)$
are given by a general Sugawara construction \cite{Moh}:
\begin{eqnarray}
T(z)&=&\!\Omega_{ab}\!:\!(\!J^aJ^b\!+\!\partial\psi^a\psi^b)\!:\!(z)
+{2\over
 3}\Omega_{ad}\Omega_{be}f^{de}_{~~c}\!
:\!(\!J^a\!:\!\psi^b\psi^c\!:+\psi^a\!:(\!\psi^bJ^c-J^b\psi^c)
\!:\!)\!:\!(z),~~
\label{eq:Tz_gen}\\
G(z)&=&\!2\Omega_{ab}:\!J^a\psi^b\!:\!(z)+{4\over
 3}\Omega_{ad}\Omega_{be}f^{de}_{~~c}\!:\!\psi^a\!:\!\psi^b\psi^c\!::\!(z)\,,
\label{eq:Gz_gen}
\end{eqnarray}
where $\Omega_{ab}$ is the inverse of $\Omega^{ab}$: 
$\Omega^{ab}\Omega_{bc}=\delta^{a}_c$.
In fact, they satisfy the following OPEs:
\begin{eqnarray}
T(y)\psi^a(z)&\sim&{1\over (y-z)^2}{1\over 2}\psi^a(z)+
{1\over y-z}\partial\psi^a(z)
\,,
\label{eq:OPEs1}\\
T(y)J^a(z)&\sim&{1\over (y-z)^2}J^a(z)+
{1\over y-z}\partial J^a(z)
\,,\\
G(y)\psi^a(z)&\sim&{1\over y-z}J^a(z)\,,
\label{eq:OPEs2}\\
G(y)J^a(z)&\sim&{1\over (y-z)^2}\psi^a(z)+{1\over y-z}\partial\psi^a(z)
\,,\\
T(y)T(z)&\sim& {c\over 2}{1\over (y-z)^4}+{1\over (y-z)^2}2T(z)
+{1\over y-z}\partial T(z)\,,\\
G(y)G(z)&\sim&{2c\over 3} {1\over (y-z)^3}+{1\over y-z}\,2T(z)\,,\\
T(y)G(z)&\sim& {1\over (y-z)^2}{3\over 2}G(z)+{1\over y-z}\partial G(z)\,,
\end{eqnarray}
and the central charge $c$ is given by $c={3\over 2}{\rm
dim}{\cal G}-f^{ac}_{~~d}f^{bd}_{~~c}\Omega_{ab}$ \cite{Moh}.

Let us consider the Berkovits' open superstring field theory
on the above CFT background.
The action has the same form as the flat one:
\begin{eqnarray}
\label{eq:actionBOZt}
 S[\Phi]&=&-{1\over g^2}\int_0^1dt
\langle\!\langle(\eta_0\Phi)(e^{-t\Phi}\Q e^{t\Phi})\rangle\!\rangle
\end{eqnarray}
 although $T(z)$ and $G(z)$ in the definition of the BRS
 operator\footnote{
The BRS operator is given by the matter Virasoro operators $T(z),G(z)$ and
ghosts $(b,c,\phi,\xi,\eta)$ as:
\begin{eqnarray}
 \Q&=&\oint{dz\over 2\pi i}\left[c\left(T-{1\over
				    2}(\partial\phi)^2
-\partial^2\phi+\partial \xi\eta
\right)\!(z)+bc\partial c(z)+\eta e^{\phi}G(z)-\eta\partial\eta e^{2\phi}b(z)
\right].
\label{eq:QBord}
\end{eqnarray}
} are given by (\ref{eq:Tz_gen}) and (\ref{eq:Gz_gen}).
The star product among string fields in the action 
is constructed by LPP's method  \cite{LPP1} in terms of conformal
mappings and correlators in the above CFT.
In order to make $\Q$ nilpotent, we assume that
the total central charge in the matter sector is $c=15$.
With this setup, we shall show that
\begin{eqnarray}
&&\Phi_0=-\tilde{V}^a_L(F_a)I\,,
\label{eq:solution_gen}\\
&&\tilde{V}^a_L(F_a)=\int_{C_{\rm left}}{dz\over 2\pi i}
F_a(z){\tilde v}^a(z),~~F_a(-1/z)=z^2F_a(z)\,,
\end{eqnarray}
is  a classical solution, where the operator $\tilde{v}^a(z)$ is
given by the lowest component of supercurrent ${\mathbf{J}}^a(z,\theta)$
with dimension $1/2$ and appropriate ghost part:
\begin{eqnarray}
\label{eq:vtila_g}
 \tilde{v}^a(z)&=&{1\over \sqrt{2}}c\xi e^{-\phi}\psi^a(z)\,,
\end{eqnarray}
and $I$ is the identity string field which is the identity element with
respect to the star product. Noting $\tilde{v}^a(z)$ is a primary field
with dimension $0$ and therefore satisfies
\begin{eqnarray}
 (\tilde{V}^a_L(F_a)I)*B=-\tilde{V}^a_R(F_a)I*B
=I*(\tilde{V}^a_L(F_a)B)=\tilde{V}^a_L(F_a)B
\end{eqnarray}
for any string field $B$ 
where $\tilde{V}^a_R(F_a)=\int_{C_{\rm right}}{dz\over 2\pi i}
F_a(z){\tilde v}^a(z)$, we obtain
\begin{eqnarray}
 e^{-\Phi_0}*\Q e^{\Phi_0}
&=&(e^{\tilde{V}^a_L(F_a)}\Q e^{-\tilde{V}^a_L(F_a)})I
=\left(-V_L^a(F_a)+{1\over 8}\Omega^{ab}C_L(F_aF_b)\right)I\,,
\label{eq:omegaI}
\end{eqnarray}
where
\begin{eqnarray}
&&V_L^a(F_a)=\int_{C_{\rm left}}{dz\over 2\pi i}
F_a(z)v^a(z)\,,~~~v^a(z)\equiv{1\over \sqrt{2}}cJ^a(z)+{1\over \sqrt{2}}
\eta e^{\phi}\psi^a(z)\,.
\end{eqnarray}
In the above computation, we have used the relations
\begin{eqnarray}
&&[\Q,\tilde{v}^a(z)]=v^a(z)\,,
\label{eq:QBvtil}
\\
&&\tilde{v}^a(y)v^b(z)\sim{1\over y-z}{-1\over 4}\Omega^{ab}c(z)\,,
~~~~~[\tilde{V}^a_L(f),V^b_L(g)]=-{1\over 4}\Omega^{ab}C_L(fg)\,,
\end{eqnarray}
which follow from
(\ref{eq:OPEs1}), (\ref{eq:OPEs2}), (\ref{eq:OPEJ1})
and (\ref{eq:OPEJ2}). Because both $V_L^a(F_a)$ and $C_L(F_aF_b)$
in (\ref{eq:omegaI}) do not include $\xi_0$, we conclude that 
$\Phi_0$ (\ref{eq:solution_gen}) satisfies the equation of motion:
\begin{eqnarray}
 \eta_0(e^{-\Phi_0}*\Q e^{\Phi_0})=0\,.
\end{eqnarray}
By replacing $F_a(z)$ with $tF_a(z)$, we have
$\eta_0(e^{-t\Phi_0}\Q e^{t\Phi_0})=0$ ($0\le t\le 1$), which implies
that the value of the action at this solution is zero: $S[\Phi_0]=0$
as we can easily check from eq.~(\ref{eq:actionBOZt}).

Around the solution $\Phi_0$,
 using (\ref{eq:omegaI}) and (\ref{eq:Q_B'formula})
and noting
\begin{eqnarray}
 A*(V_L^a(F_a)I)=-(-1)^{|A|}(V_R^a(F_a)A)*I=-(-1)^{|A|}V_R^a(F_a)A\,,
\end{eqnarray}
($V^a_R(f)=\int_{C_{\rm right}}\!{dz\over 2\pi i}
F_a(z)v^a(z)$) for any string field $A$ 
because $v^a(z)$ is a primary field with dimension $0$,
the new BRS operator $\Q'$ which is a derivation with respect to the star
product becomes
\begin{eqnarray}
&&\Q'B\nonumber\\
&&=\Q B+\!\left[\!\left(\!-V_L^a(F_a)+{1\over
	       8}\Omega^{ab}C_L(F_aF_b)\!\right)\!I\right]\!*\!B
-\!(-1)^{|B|}\!B\!*\!\left[\!\left(\!-V_L^a(F_a)+{1\over
	      8}\Omega^{ab}C_L(F_aF_b)\!\right)\! I\right]
\nonumber\\
&&=\left(\Q
-(V^a_L(F_a)+V_R^a(F_a))+{1\over 8}\Omega^{ab}(C_L(F_aF_b)+C_R(F_aF_b))
\right)B\,,
\end{eqnarray}
on any string field $B$, namely, 
\begin{eqnarray}
&&\Q'=\Q-V^a(F_a)+{1\over 8}\Omega^{ab}C(F_aF_b)\,,
\label{eq:QBp_gen}\\
&&V^a(F_a)=\oint{dz\over 2\pi i}F_a(z)v^a(z)\,,~~~
C(F_aF_b)=\oint{dz\over 2\pi i}F_a(z)F_b(z)c(z)\,.
\end{eqnarray}
We can directly check $\{\eta_0,\Q'\}=0$ and nilpotency $\Q^{\prime
2}=0$ noting $\{\Q,v^a(z)\}=0$ from (\ref{eq:QBvtil}) and 
\begin{eqnarray}
 &&v^a(y)v^b(z)\!\sim\!{-\Omega^{ab}\over 4(y-z)}(c\partial c
-\eta\partial\eta e^{2\phi})(z),~~
\{V^a(F_a),V^b(F_b)\}={-\Omega^{ab}\over 4}\{\Q,\!C(F_aF_b)\}.~~~~~~~
\end{eqnarray}
We note that the above BRS operator $\Q'$ is obtained by replacing
the matter Virasoro operators $G(z),T(z)$ in (\ref{eq:QBord}) with
$G(z)-{1\over \sqrt{2}}F_a(z)\psi^a(z)$,  
$T(z)-{1\over \sqrt{2}}F_a(z)J^a(z)+{1\over
 8}\Omega^{ab}F_a(z)F_b(z)$ 
respectively. In fact, if we define $G'(z)=\sum_{r}G'_rz^{-r-3/2}$ 
and $T'(z)=\sum_{n}L_n'z^{-n-2}$  as
\begin{eqnarray}
\label{eq:Gp}
G'_r&=&G_r-{1\over \sqrt{2}}\sum_{k}F_{a,k}\psi^a_{r-k}\,,\\
 L'_n&=&L_n-{1\over \sqrt{2}}\sum_{k}F_{a,k}J^a_{n-k}+{1\over 8}\Omega^{ab}
\sum_{k}F_{a,n-k}F_{b,k}\,,
\label{eq:Lp}
\end{eqnarray}
where $\psi^a(z)=\sum_r\psi_r^az^{-r-1/2},J^a(z)=\sum_nJ^a_nz^{-n-1},
G(z)=\sum_{r}G_rz^{-r-3/2},T(z)=\sum_{n}L_nz^{-n-2}$ and
$F_{a,n}=\oint {d\sigma\over 2\pi}e^{i(n+1)\sigma}F_a(e^{i\sigma})$,
then, using OPEs among $(\psi^a,J^a,G,T)$,
we can check that they satisfy the super Virasoro algebra:
\begin{eqnarray}
&&[L_m',L_n']=(m-n)L_{m+n}'+{c\over 12}(m^3-m)\delta_{m+n,0}\,,\\
&&\{G'_r,G'_s\}=2L'_{r+s}+{c\over 12}(4r^2-1)\delta_{r+s,0}\,,\\
&&[L_m',G_r']=\left({m\over 2}-r\right)G_{m+r}'\,,
\end{eqnarray}
with the same central charge as original $G(z),T(z)$ system.
Furthermore, let us define $\psi^{\prime a}(z)=\sum_r\psi^{\prime
a}_rz^{-r-1/2}$
and $J^{\prime a}(z)=\sum_nJ^{\prime a}_nz^{-n-1}$ by
\begin{eqnarray}
&&\psi^{\prime a}_r=\sum_kM^a_{~b,k}\psi^b_{r-k}\,,~~~~
J^{\prime a}_n=\sum_kM^a_{~b,k}\left(J^b_{n-k}-{1\over 2\sqrt{2}}
\Omega^{bc}F_{c,n-k}\right)\,,
\label{eq:currentp}
\end{eqnarray}
where $M^a_{~b,n}$ is given by a path-ordered form:
\begin{eqnarray}
 M^a_{~b}(\sigma)&=&\sum_{n}M^a_{~b,n}e^{-in\sigma}=
\left[{\mathbf{P}}\exp\left(i\int_0^1dt\, \sigma
A(t\sigma)\right)\right]^a_{~b}
\nonumber\\
&=&\!\delta^a_b\!+\!\sum_{n=1}^{\infty}i^n\sigma^n\!\int_0^1\!dt_1
\int_0^{t_1}\!dt_2
\cdots 
\int_0^{t_{n-1}}\!dt_n
 A^a_{~c_n}\!(t_n\sigma)
A^{c_n}_{~c_{n-1}}\!(t_{n-1}\sigma)
\cdots 
A^{c_2}_{~b}(t_1\sigma),~~~~\label{eq:path_o}
\\
A^a_{~b}(\sigma)
&\equiv&{1\over \sqrt{2}}f^{ac}_{~~b}e^{i\sigma}F_c(e^{i\sigma})\,.
\end{eqnarray}
Noting the identities for invariant metric $\Omega^{ab}$
(\ref{eq:Oinvariance}) and the Jacobi identity for structure constants
$f^{ab}_{~~c}$, we can show following relations:
\begin{eqnarray}
&& -i\partial_{\sigma}M^a_{~b}(\sigma)=M^a_{~c}(\sigma)
A^c_{~b}(\sigma)\,,~~~~~~~
-nM^a_{~b,n}={1\over \sqrt{2}}\sum_kM^a_{~d,k}f^{dc}_{~~b}F_{c,n-k}\,,
\label{eq:formula_M1}\\
&&M^a_{~c}(\sigma)M^b_{~d}(\sigma)\Omega^{cd}=\Omega^{ab}\,,~~~~~~~~~~~~
\sum_{k}M^a_{~c,k}M^b_{~d,n-k}\Omega^{cd}=\Omega^{ab}\delta_{n,0}\,,
\label{eq:formula_M2}\\
&&M^a_{~d}(\sigma)M^b_{~e}(\sigma)f^{de}_{~~c}
=f^{ab}_{~~d}M^d_{~c}(\sigma)\,,~~~~
\sum_kM^a_{~d,k}M^b_{~e,n-k}f^{de}_{~~c}=f^{ab}_{~~d}M^d_{~c,n}\,,
\label{eq:formula_M3}
\end{eqnarray}
and we obtain commutation relations:
\begin{eqnarray}
 &&\{\psi^{\prime a}_r,\psi^{\prime b}_s\}
={1\over 2}\Omega^{ab}\delta_{r+s,0},
~[J^{\prime a}_m,J^{\prime b}_n]={1\over
2}\Omega^{ab}m\delta_{m+n,0}+f^{ab}_{~~c}J^{\prime c}_{m+n},
~[J^{\prime a}_n,\psi^{\prime b}_r]=f^{ab}_{~~c}\psi^{\prime
c}_{n+r},~~~~~~~\\
&&[L_n',\psi^{\prime a}_r]=-\left({n\over 2}+r\right)\psi^{\prime
 a}_{n+r}\,,~
~~~[L_m',J^{\prime a}_n]=-nJ^{\prime a}_{m+n},\\
&&\{G_r',\psi_s^{\prime
a}\}=J^{\prime a}_{r+s}\,,~
~~~[G_r',J^{\prime a}_n]=-n\psi^{\prime a}_{n+r},
\end{eqnarray}
which are the same form as the original (unprimed) ones.

After all, by re-expanding the action (\ref{eq:actionBOZt})
around a classical solution $\Phi_0$ (\ref{eq:solution_gen})
as $e^{\Phi}=e^{\Phi_0}e^{\Phi'}$,
we obtain the action $S'[\Phi']$ with new BRS operator
(\ref{eq:QBp_gen}), which is realized by 
a replacement $(\psi^a,J^a,G,T)\rightarrow (\psi^{\prime a},J^{\prime
a},G',T')$ in eqs.~(\ref{eq:Gp}), (\ref{eq:Lp}) and (\ref{eq:currentp})
preserving algebra among them.
This fact and vanishing vacuum energy: $S[\Phi_0]=0$
suggest that the classical solution $\Phi_0$ (\ref{eq:solution_gen})
might be a pure gauge solution in terms of superstring field theory.
Indeed, the new BRS operator (\ref{eq:QBp_gen}) can be rewritten as
a similarity transform from the original one:
\begin{eqnarray}
\label{eq:similarity_gen}
 \Q'=e^{\tilde{V}^a(F_a)}\Q e^{-\tilde{V}^a(F_a)}\,,~~~~~
\tilde{V}^a(F_a)=\oint{dz\over 2\pi i}F_a(z){\tilde v}^a(z)\,.
\end{eqnarray}
One might think that the original action $S[\Phi'']$
could be reproduced by a field redefinition such as
$\Phi''=e^{-\tilde{V}^a(F_a)}\Phi'
=e^{-\tilde{V}^a_L(F_a)I}*\Phi'*e^{\tilde{V}_L^a(F_a)I}$
in the re-expanded action $S'[\Phi']$.
However, it is not so trivial because there is another derivation
$\eta_0$ in the action and $[\eta_0,\tilde{V}^a(F_a)]\ne 0$.

In the following, we demonstrate that if the function $F_a(z)$ satisfies
a condition, we can explicitly rewrite our solution in a pure gauge form
and take an appropriate field redefinition around it,
which recovers original action.
Let us consider a particular pure gauge form and try to rewrite
our solution to it.
Noting the commutation relation $[\Q,J^a(z)-\partial(c\xi
e^{-\phi}\psi^a)(z)]=0$, we find an identity:
\begin{eqnarray}
 J^a(z)&=&\partial(c\xi e^{-\phi}\psi^a)(z)+\{\Q,\Omega^a(z)\}\,,\\
\Omega^a(z)&\equiv&{1\over 2}
c\partial c\xi\partial \xi \partial^2\xi e^{-3\phi}\psi^a(z)
-c\xi\partial\xi e^{-2\phi}J^a(z)\,,
\end{eqnarray}
which relates  $\tilde{v}^a$ (\ref{eq:vtila_g})
to the current $J^a$ and we have
\begin{eqnarray}
\label{eq:JvQ}
&&\tilde{V}^a_L(\partial g_a)+J^a_L(g_a)=\{\Q,\Omega^a_L(g_a)\}\,,\\
&&J_L^a(g_a)=\int_{C_{\rm left}}{dz\over 2\pi i}{1\over \sqrt{2}}
g_a(z)J^a(z),~~
\Omega^a_L(g_a)=\int_{C_{\rm left}}{dz\over 2\pi i}{1\over \sqrt{2}}
g_a(z)\Omega^a(z),~~g_a(\pm i)=0\,.~~~~~~~~
\end{eqnarray}
The last condition is necessary to 
remove unwanted boundary contribution of the integration
given by  $g_a(\pm i)\tilde{v}^a(\pm i)$.
We notice commutation relations
\begin{eqnarray}
~[\tilde{V}_L^a(F_a),\tilde{V}_L^b(G_b)]=0,~~~[\tilde{V}_L^a(F_a),J_L^b(g_b)]
={1\over \sqrt{2}}f^{ab}_{~~c}\tilde{V}_L^c(F_ag_b),
\label{eq:comm2}
\end{eqnarray}
and a kind of Hausdorff formula:
\begin{eqnarray}
 e^{A}e^{B}&=&\exp\left(A+{\rm ad}_{A\over 2}(1+\coth({\rm ad}_{A\over
		   2}))B+{\rm O}(B^2)\right),
\end{eqnarray}
where we have denoted ${\rm ad}_XY=[X,Y]$
and ${\rm O}(B^2)$ is quadratic and higher terms with respect to $B$.
By substituting $A=\tilde{V}^a_L(\partial
g_a)+J^a_L(g_a)$ and $B=-\tilde{V}_L^a(F_a)$ to the above formula,
we obtain
\begin{eqnarray}
e^{\tilde{V}^a_L(\partial
g_a)+J^a_L(g_a)}e^{-\tilde{V}_L^a(F_a)}&=&e^{J^a_L(g_a)
+\tilde{V}^a_L(\partial g_a-F_b({\cal M}(e^{{\cal
M}}-1)^{-1})^b_{~a})}\,,\\
{\cal M}^b_{~a}(z)&=&{1\over \sqrt{2}}f^{bc}_{~~a}g_c(z)\,,
\end{eqnarray}
because ${\rm O}(B^2)$ in the exponent vanishes due to
(\ref{eq:comm2}).  If the second term in the exponent on the right
hand side of the first line vanishes, we can compute as:
\begin{eqnarray}
 &&e^{\Q\Omega^a_L(g_a)I}e^{-\tilde{V}_L^a(F_a)I}=e^{(\tilde{V}^a_L(\partial
  g_a)+J^a_L(g_a))}e^{-\tilde{V}_L^a(F_a)}I
=e^{J_L^a(g_a)}I=e^{\eta_0\xi_0J_L^a(g_a)I},
\end{eqnarray}
where we should impose $g_a(-1/z)=g_a(z)$ 
to guarantee a relation such as eq.~(\ref{Eq:LRstar}).
The above calculation means that, 
by solving a differential equation with respect to $g_a(z)$:
\begin{eqnarray}
\label{eq:pure_gen}
&&F_b(z)=\partial g_a(z)((e^{{\cal M}}-1){\cal M}^{-1})^a_{~b}(z),\\
&&g_a(-1/z)=g_a(z),~~~g_a(\pm i)=0,
\label{eq:cond_g}
\end{eqnarray}
for a given $F_a(z)$ which specifies the classical solution $\Phi_0$
(\ref{eq:solution_gen}), we obtain the form of gauge 
transformation from the trivial solution $\Phi=0$
in superstring field theory:
\begin{eqnarray}
\label{eq:pure_g_gen}
 e^{\Phi_0}&=&e^{\Q(-\Omega^a_L(g_a))I}e^{\eta_0(\xi_0 J_L^a(g_a))I}\,.
\end{eqnarray}
Note that the equation (\ref{eq:pure_gen}) is consistent with the
conditions $F_a(-1/z)=z^2F_a(z)$ and $g_a(-1/z)=g_a(z)$.
For an abelian ${\cal G}$, where ${\cal M}^b_{~a}(z)=0$,
 it  becomes a simple form $\partial g_a(z)=F_a(z)$
because of $((e^{{\cal M}}-1){\cal M}^{-1})^a_{~b}(z)
=\delta^a_b+{\rm O}({\cal M})$.
Similarly, we can rewrite the new BRS operator (\ref{eq:QBp_gen}) as
\begin{eqnarray}
\label{eq:sim_global}
 \Q'&=&e^{-\Lambda}\Q e^{\Lambda},~~~~
\Lambda=-\oint{dz\over 2\pi i}{1\over \sqrt{2}}g_a(z)J^a(z)
\end{eqnarray}
using the above $g_a$. Thanks to $[\eta_0,\Lambda]=0$, we can recover
the original action by taking a field redefinition
$\Phi''=e^{\Lambda}\Phi'=
e^{{1\over 2}\lambda_a\oint{dz\over 2\pi}J^a(z)+\cdots}
\Phi'$, which corresponds to a marginal deformation 
by the current $J^a$ \cite{RS}. Its deformation parameter 
$\lambda_a=\sqrt{2} i\int_{-\pi}^{\pi}{d\sigma\over
2\pi}g_a(e^{i\sigma})$ is related to the function $F_a$ in
our classical solution by eq.~(\ref{eq:pure_gen}).
However, a solution to the differential equation  (\ref{eq:pure_gen})
which satisfies the conditions (\ref{eq:cond_g}) does not necessarily
exist.
In fact, eq.~(\ref{eq:pure_gen}) can be rewritten as,
\begin{eqnarray}
\label{eq:diff_non}
 {1\over \sqrt{2}}F_b(z)T^b&=&e^{-{1\over \sqrt{2}}g_a(z)T^a}
\partial\left(e^{{1\over \sqrt{2}}g_a(z)T^a}\right),
\end{eqnarray}
where $T^a$s are generator matrices of ${\cal G}$ 
such as $[T^a,T^b]=f^{ab}_{~~c}T^c$
and use has been made of 
$(e^{{\cal M}}-1){\cal M}^{-1}=\int_0^1 dt e^{t{\cal M}}$ and 
$\delta (e^X) = \int_0^1 dt e^{(1-t) X}\delta X e^{tX}$.
This equation can be solved by path-ordered form:
\begin{eqnarray}
e^{{1\over \sqrt{2}}g_a(e^{i\sigma})T^a}
&=&\mathbf{P}\,e^{
{i\over \sqrt{2}}
\int_0^1dt\left(\sigma-{\rm sgn}(\sigma){\pi\over 2}\right)
e^{it\sigma+i{\rm sgn}(\sigma){\pi\over 2}(1-t)}
F_a(e^{it\sigma+i{\rm sgn}(\sigma){\pi\over 2}(1-t)})T^a
},
\end{eqnarray}
where the path ordering denoted by $\mathbf{P}$ is taken as 
(\ref{eq:path_o}) (i.e., we put a 
matrix associated with larger $t$ to the right)
and ${\rm sgn}(\sigma)=+1\,(-1)$ for $\sigma>0\,(\sigma<0)$ is used.
Here we have respected the second condition in (\ref{eq:cond_g}): 
$g_a(\pm i)=0$ and solved separately on upper and lower half circle
by taking the phase of $z=e^{i\sigma}$ as $-\pi\le \sigma \le \pi$
on the unit circle.
The property of the function $F_a(z)$ in our
solution $\Phi_0$ (\ref{eq:solution_gen}): $F_a(-1/z)=z^2F_a(z)$ implies 
$e^{i({\rm sgn}(\sigma)\pi-\sigma)}F_a(e^{i({\rm
sgn}(\sigma)\pi-\sigma)})=-e^{i\sigma}F_a(e^{i\sigma})$ and then
the above solution satisfies the first condition 
in (\ref{eq:cond_g}): $g_a(e^{i({\rm sgn}(\sigma)\pi-\sigma)})
=g_a(e^{i\sigma})$.
In order to guarantee the continuity of $g_a(e^{i\sigma})$  at
$\sigma=0$, which is needed for (\ref{eq:JvQ}),
there is a consistency condition for $F_a(z)$:
\begin{eqnarray}
\label{eq:wilson_non}
\mathbf{P}\,e^{
{-i\pi \over 2\sqrt{2}}
\int_0^1dt e^{i{\pi\over 2}(1-t)}
F_a(e^{i{\pi\over 2}(1-t)})T^a}
=
\mathbf{P}\,e^{
{i\pi \over 2\sqrt{2}}
\int_0^1dt e^{-i{\pi\over 2}(1-t)}
F_a(e^{-i{\pi\over 2}(1-t)})T^a},
\end{eqnarray}
which is reduced to 
\begin{eqnarray}
\label{eq:Wilson_l_gen}
 \int_{C_{\rm left}}{dz}F_a(z)=0
\end{eqnarray}
in the case of ${\cal G}$ : abelian.
Namely, if $F_a(z)$ satisfies the condition (\ref{eq:wilson_non}) (or
(\ref{eq:Wilson_l_gen}) for abelian ${\cal G}$), 
the solution (\ref{eq:solution_gen}) is rewritten 
in a pure gauge form (\ref{eq:pure_g_gen})
and induces a field redefinition generated by $\Lambda$ in
(\ref{eq:sim_global}).
Conversely, in the case that $F_a(z)$
breaks the condition (\ref{eq:wilson_non}),
we cannot 
rewrite as (\ref{eq:pure_g_gen})
and we should know further informations about
the supercurrent and its representation by specifying a model
which realizes ${\bf J}^a(z,\theta)$ in order to find
explicit relations between our solution and marginal deformation.

As an example of ${\cal G}=u(1)^{10}$, we take an ordinary flat
background which is described by a supercurrent 
$\mathbf{J}^{\mu}(z,\theta)=\psi^{\mu}(z)+
\theta {i\over \sqrt{2\alpha'}}\partial X^{\mu}(z)$.
In this case, we can identify various quantities as follows:
\begin{eqnarray}
&&T(z)=-{1\over4 \alpha'}\partial X^{\mu}
\partial X_{\mu}(z)-{1\over 2}\psi^{\mu}\partial\psi_{\mu}(z)\,,
~~~G(z)= {i\over \sqrt{2\alpha'}}\partial X_{\mu}\psi^{\mu}(z)\,,
~~~~\\
&&\Omega^{\mu\nu}=2\eta^{\mu\nu}\,,~~~
\Omega_{\mu\nu}={1\over 2}\eta_{\mu\nu}\,,~~~f^{\mu\nu}_{~~\rho}=0\,,
~~~~c={3\over 2}{\rm dim}(u(1)^{10})=15\,.
\end{eqnarray}
By taking functions $F_{\mu}(z)$ for a solution $\Phi_0$
(\ref{eq:solution_gen}) as $F_{\mu}(z)=\delta_{\mu,9}F(z)$
such as $F(-1/z)=z^2F(z)$,
we reproduce the solution in the previous section,
which has turned out to correspond to the Wilson line.
From the condition (\ref{eq:Wilson_l_gen}), non-vanishing Wilson line
$f=\int_{C_{\rm left}}{dz\over 2\pi i}F(z)\ne 0$ implies non-existence
of a function $g_{\mu}(z)$ which specifies globally defined gauge parameter of
the form (\ref{eq:pure_g_gen}) and a field redefinition associated with
(\ref{eq:sim_global}).
Instead, we have found other (local) expressions (\ref{eq:local_pure})
and (\ref{Eq:redef}) using the integration of the current $J^{\mu}(z)$.

We comment on the analogy with arguments in
the Witten's bosonic string field theory (see appendix
\ref{sec:bosonic}).
In both cases, we can construct a class of classical solutions
(\ref{eq:solution_gen}), (\ref{eq:solution_b})
based on (super-)currents, which have vanishing vacuum energy.
The actions around the solution are also obtained by
appropriate operator mappings (\ref{eq:currentp}), (\ref{eq:mapj_b})
which preserve current algebra in both cases.
This fact suggests that the theory around the 
solutions is essentially the same as the original one
and they might be gauge equivalent.
Indeed, we can represent the solutions as a pure gauge form in both 
supersymmetric and bosonic string field theory
if there exists a solution $g_a(z)$ to
the differential equation~(\ref{eq:pure_gen}), which is the same form
as (\ref{eq:pure_eqb}).
In both cases, the change of BRS operator around the
solution turns out to be absorbed by a field redefinition
using $g_a(z)$.
As was shown in the previous section
for supersymmetric case and in ref.~\cite{rf:TT1} for bosonic case,
 we can obtain nontrivial solutions in a global sense
by considering a compactified background with the non-vanishing Wilson lines
$\int_{C_{\rm left}} dz F_{\mu}(z)\ne 0$.
In general case, there is a possibility that our classical solutions become
nontrivial if $F_a(z)$ breaks the condition (\ref{eq:wilson_non}),
although we cannot prove their non-triviality with respect to gauge
transformation of string field theory at this stage
because we have only investigated a particular pure gauge form:
(\ref{eq:pure_g_gen}) or ({\ref{eq:pure_gen_b}}).
We speculate that they will be rewritten as a {\it locally} pure gauge
form using a kind of integration of the current
in each model in the case that $g_a(z)$ does not exist.

%%%%%%%%%%%%%%%%%%%%%%%%%%%%%%%%%%%%%%%%%%%%%%%%%%%%%%%%%%%%%%%%
\section{Discussions}

We constructed a class of analytic classical solutions in open
superstring field theory, which is related to marginal deformations in
conformal field theory. We showed that the resulting solutions can be
represented using a well-defined Fock space expression, and the vacuum
energy vanishes due to the ghost number non-conservation in the large
Hilbert space. For the solution corresponding to background Wilson
lines, we observe that the solution can be written as a locally pure
gauge form, and the action expanded around the solution can be
transformed locally back to the original action by a string field
redefinition. 
The analytic classical solution enabled us to investigate gauge
structure in the string field theory. We found that the half integration
mode of the function in the solution is unchanged under the ``global
transformation'', but other modes can be gauged away. 
Space-time supersymmetry is realized on-shell in the theory and the
solution is a supersymmetric solution.
We note that the classical solutions in the present paper can be easily  
extended to that of the theory given in ref.~\cite{rf:BSZ} including the
GSO$(-)$ sector  

We have extensively used formal properties of the identity string field 
to construct our solution and to investigate its structure.
In general, the identity string field requires careful handling 
to evaluate some quantities of the form $\langle\!\langle ({\cal O}_1I) 
({\cal O}_2 I)\rangle\!\rangle$. Actually, in terms of oscillator
representation, 
we encounter divergence from contractions of nonzero modes
in computing $\langle I|\cdots |I\rangle$ as in the case of bosonic
string field theory. Consequently, it is necessary to 
define appropriate regularization of the identity string field, which is
not yet known,
in both super and bosonic string field theory.
However, we evaluated the vacuum energy
at our solution $\Phi_0$ using the
zero-mode saturation rule for the $\xi\eta$ ghost system
in the large Hilbert space.
We hope to obtain information about a consistent regularization
by comparing these calculations.
The situation is different in bosonic string field theory, where
we gave a formal proof (\ref{eq:vace_b}) to show 
that the vacuum energy at the solution corresponding to the
marginal deformation vanishes as in refs.~\cite{rf:TT2,rf:KugoZwie}.
It is preferable to prove it more directly.

We briefly comment on the modified version of {\it cubic} superstring
field theory \cite{rf:PTY,rf:Arefeva}.
Setting the Ramond field to zero, the equation of motion is given by
$Y_{-2}(\Q A+A*A)=0$,
where $Y_{-2}$ is the picture changing operator with picture number
$(-2)$.
We note that a string field $A_0=e^{-\Phi_0}\Q e^{\Phi_0}$,
in which  $\Phi_0$  is our solution 
in the Berkovits' theory,
is a solution to the equation of motion. In fact,
$\Q A_0+A_0*A_0=0$ holds
and $A_0$ itself has ghost number $1$ and picture number
 $0$ and is Grassmann odd in the {\it small} Hilbert space.
Therefore, we find that $A_0$ is pure gauge with the gauge parameter
$J_L^a(g_a)I$ if $F_a(z)$ in $\Phi_0$ satisfies the condition
(\ref{eq:wilson_non}). 
We notice that 
$\Phi_0$ itself cannot be a gauge parameter in spite of the form of $A_0$
 because $\Phi_0$ is {\it not} in the small Hilbert space :
 $\eta_0\Phi_0\ne 0$.

For the Wilson line solution, the half integration of the function is
unchanged under the global transformation.
This is an obvious result because the half integration mode as the
Wilson line should be a physical observable.
However, we confirmed the invariance of the half
integration mode merely for a part of the gauge symmetry, and it is
difficult to prove the invariance for the whole gauge symmetry.
More precisely, we have to relate the solution to a general gauge
invariant quantity in string field theory. Although a gauge invariant
quantity plays important 
roles in field theories, we have not yet understood completely how it
can be constructed in string field theory. As gauge invariants, we know 
the action and some operators only \cite{InterpltSFT,Michi2}.
It is natural to ask how the Wilson loop operator is generalized
in string field theory. This is an important open question.

In the theory expanded around the classical solution, the background of
the theory can be changed from the unexpanded theory. 
If we choose $su(2)$ currents, the solution corresponds to the tachyon
lump solution \cite{rf:TT1,rf:KTZ} (see also appendix D).
In the background, the boundary condition of a string coordinate is
changed from the Neumann one to the Dirichlet one \cite{rf:Sen1}.
In string field theory, we found this phenomenon indirectly by studying
a gauge invariant operator \cite{rf:KTZ}.
On the other hand, in ref.~\cite{rf:KugoZwie}, it was proposed that
string coordinates 
$X^\mu(\sigma)$ and its conjugate momenta $P_\mu(\sigma)$ are
universal objects in string field theory, and the various backgrounds
correspond to inequivalent representations of their canonical algebra.
In the present case, it seems that the tachyon lump solution changes
representation of universal coordinates, namely their boundary
conditions.
This subject was studied from the viewpoint of vacuum string
field theory \cite{rf:HM}. 

It is an important problem to construct an analytic solution
representing tachyon condensation in the superstring field theory. 
In order to investigate this subject, we have to 
introduce Chan-Paton matrices to include both GSO$(+)$ and GSO$(-)$
sectors. Such a formulation were developed in ref.~\cite{rf:BSZ}.
We hope to obtain an analytic expression
of the tachyon vacuum by an analogous construction to the bosonic case
(i.e., a class of scalar solutions in refs.~\cite{rf:TT2,rf:KT}).

%%%%%%%%%%%%%%%%%%%%%%%%%%%%%%%%%%%%%%%%%%%%%%%%%%%%%%%%%%%%%%%%
\section*{Acknowledgements}

The authors would like to thank Yuji Igarashi and Katsumi Itoh for
useful discussions.
I.~K. wishes to express his gratitude to Kazuki Ohmori for valuable
comments.

\appendix
%%%%%%%%%%%%%%%%%%%%%%%%%%%%%%%%%%%%%%%%%%%%%%%%%%%%%%%%%%%%%%%%
\section{Oscillator expression of the identity string field
\label{sec:Osc}}

We use the identity string field in constructing 
exact solutions in the framework of Berkovits' open superstring field
theory. The action is described in terms of the {\it large} Hilbert
space \cite{FMS} in the NSR formalism.
We shall give an explicit oscillator representation of the identity
string field $|I\rangle$ in terms of modes of
$X^{\mu},\psi^{\mu},b,c,\phi,\xi,\eta$ in the NS sector.
In this paper, we formally regard the identity string field $I$ as the
identity element with respect to the Witten $*$ product: $A*I=I*A=A$,
which should be proved by LPP's definition of string vertices
\cite{LPP1} and generalized gluing and resmoothing theorem 
at the critical dimension $d=10$ \cite{LPP2, SS, KO}.
Therefore, we define $\langle I|$ as a 1-string LPP vertex 
using conformal mapping: $h_I(z)=2z/(1-z^2)$ \cite{RZ}
and CFT correlator in the large Hilbert space
denoted by $\langle\!\langle \cdots \rangle\!\rangle$
which is evaluated on the upper half plane:\footnote{
In this paper, we implicitly use the doubling trick: a holomorphic field
$\sigma(z)$ and
antiholomorphic one $\tilde{\sigma}(\bar{z})$ 
in the upper half plane are combined into a holomorphic field defined in
the whole complex plane with a boundary condition
$\sigma(z)=\tilde{\sigma}(\bar{z})$ on the real axis: ${\rm Im}z=0$.
}
 $\langle I|A\rangle=\langle\!\langle h_I[{\cal
O}_A(0)]\rangle\!\rangle$, where $|A\rangle={\cal O}_A(0)|0\rangle$
is an arbitrary state. In ref.~\cite{LPP1}, the integral expression for
Neumann coefficients in $X^{\mu},b,c$ sector is given with this
definition and the oscillator expression of the identity string filed 
is obtained by applying the conformal map $h_I(z)$,  which is
consistent with that in \cite{rf:GJ1, GJ2}. The result is
\begin{eqnarray}
\label{eq:IXbc}
|I\rangle_b&=&e^{E_{Xbc}}|p^{\mu}=0\rangle\,,\\
E_{Xbc}&=&
\sum_{n\ge 1}{-(-1)^n\over 2 n}\alpha_{-n}^{\mu}
\alpha_{-n\mu}+
\sum_{n\ge 2}(-1)^nc_{-n}b_{-n}-\sum_{k\ge
 1}(-1)^k(2c_0b_{-2k}+(c_1-c_{-1})b_{-2k-1}),\nonumber
\end{eqnarray}
which is the same as the identity string field in the Witten's bosonic open
string field theory  if $\mu$ runs over $0,1,\cdots ,25$. 

In the same way, we can calculate the Neumann coefficients in the
matter fermion sector ($\psi^{\mu}$) as:
\begin{eqnarray}
I_{rs}&=&-I_{sr}
=\oint_0{dy\over 2\pi i}y^{-r-{1\over 2}}
\oint_0{dz\over 2\pi i}z^{-s-{1\over 2}}\,{(h'_I(y))^{1\over 2}
(h'_I(z))^{1\over 2}\over h_I(y)-h_I(z)}
\nonumber
\\
&=&\left\{
\begin{array}[tb]{cc}
- {r(2s-1)\over r^2-s^2}\!\left(-1\over 4\right)^{r+s\over 2}\!
{\left(r-{1\over 2}\right)!\left(s-{3\over 2}\right)! 
\over \left[\left({1\over 2}\left(r-{1\over 2}\right)\!\right)!
\left({1\over 2}\left(s-{3\over 2}\right)\!\right)!
\right]^2}
& ~~~(r-{1\over 2}:{\rm even};s-{1\over 2}:{\rm odd})\\
-{s(2r-1)\over r^2-s^2}
\left(-1\over 4\right)^{r+s\over 2}\!
{\left(r-{3\over 2}\right)!\left(s-{1\over 2}\right)! 
\over \left[\left({1\over 2}\left(r-{3\over 2}\right)\!\right)!
\left({1\over 2}\left(s-{1\over 2}\right)\!\right)!
\right]^2}
&~~~(r-{1\over 2}:{\rm odd};s-{1\over 2}:{\rm even})\\
0&({\rm otherwise})\\
\end{array}
\right.,
\end{eqnarray}
where we have used the expansion:
\begin{eqnarray}
&&\sum_{r,s\ge{1\over 2}}
(r^2-s^2)I_{rs}y^{r-{1\over 2}}z^{s-{1\over 2}}=
(y\partial_y-z\partial_z)
(y\partial_y+z\partial_z+1)\left({(h_I'(y))^{1\over 2}(h_I'(z))^{1\over
			    2}
\over h_I(y)-h_I(z)}-{1\over y-z}\right)\nonumber\\
&&=
{y(1-z^2)+z(1-y^2)\over (1+y^2)^{3\over 2}(1+z^2)^{3\over 2}}
=\!\sum_{k,l=0}^{\infty}\!(2k+1)(4l+1)\!\left(
\begin{array}[tb]{c}
 {-1\over 2}\\
k
\end{array}
\right)\!
\left(
\begin{array}[tb]{c}
 {-1\over 2}\\
l
\end{array}
\right)\!(y^{2k+1}z^{2l}+y^{2l}z^{2k+1}).~~~~~~
\end{eqnarray}
$\left(
\begin{array}[tb]{c}
 a\\
b
\end{array}
\right)={\Gamma(a+1)\over \Gamma(b+1)\Gamma(a-b+1)}$ is the binomial
coefficient.
This formula for the coefficients $I_{rs}$ is consistent with that in
\cite{GJIII}.

As for the $\phi$ sector, the formula for the Neumann coefficients is
slightly different from that of $X^{\mu}$ because of the background
charge $Q=2$ \cite{LPP2}. One can compute explicitly by substituting
$h_I(z)$ into the integrand:
\begin{eqnarray}
{\cal N}_{mn}&=&{1\over mn}\oint_0\!{dy\over 2\pi i}y^{-m}
\oint_0\!{dz\over 2\pi i}z^{-n}{h_I'(y)h_I'(z)\over 
(h_I(y)-h_I(z))^2}=-{(-1)^m\over m}\delta_{m,n}\,,
~~(m,n\ge 1),~~~\\
{\cal N}_{0n}&=&-{1\over 2n}\oint_0{dw\over 2\pi i}w^{-n}
\partial_w\log(\partial_y\partial_w\log(h_I(y)-h_I(w))|_{y=0})\nonumber\\
&=&
\left\{\begin{array}[tb]{cc}
	  {(-1)^{n\over 2}\over n}&(n:{\rm even}) \\
0&(n:{\rm odd}) 
		\end{array}
\right.,~~~~~~~(n\ge 1),\\
{\cal N}_{00}&=&0\,,
\end{eqnarray}

In the $\xi\eta$ sector, the Neumann coefficients for the identity string
field  are computed as\footnote{
In general, $N$-string vertex is given by
\begin{eqnarray}
\langle V_N|&=&\prod_r {}_r\!\langle 0|\xi^r_0\, e^{\sum_{r,s=1}^N
\sum_{n\ge 0,m\ge 1}\eta_n^rN^{rs}_{nm}\xi^s_m}(\eta_0^1+\cdots +\eta_0^N)\,,
\end{eqnarray}
in this sector, which is obtained using the method in \cite{LPP1}.
}
\begin{eqnarray}
&&N_{mn}=\oint_0\!{dy\over 2\pi i}y^{-m-1}
\oint_0\!{dz\over 2\pi i}z^{-n}{-h_I'(z)\over 
h_I(y)-h_I(z)}=-(-1)^m\delta_{m,n}\,,
~~m,n\ge 1,~~~\\
&&N_{0n}=\oint_0\!{dy\over 2\pi i}y^{-1}
\oint_0\!{dz\over 2\pi i}z^{-n}{-h_I'(z)\over 
h_I(y)-h_I(z)}=\left\{\begin{array}[tb]{cc}
	  2&~~(n:{\rm even}) \\
0&~~(n:{\rm odd}) 
		\end{array}
\right.\,,~~~n\ge 1\,,
\end{eqnarray}
where use has been made of the expansion
\begin{eqnarray}
&&{h_I'(z)\over 
h_I(y)-h_I(z)}-{1\over y-z}=-{y(1+z^2)+2z\over (1-z^2)(1+yz)}
=-\sum_{k=0}^{\infty}(2z^{2k+1}+(-1)^ky^{k+1}z^k)\,.
\end{eqnarray}

After all, the identity string field of the fermionic sector
($\psi^{\mu},\phi,\xi,\eta$) is given by
\begin{eqnarray}
 |I\rangle_f &=&e^{E_{\psi\phi\xi\eta}}|q=0\rangle\,,
\label{eq:Ipsiphi}
\\
E_{\psi\phi\xi\eta}&=&\sum_{r,s\ge
 1/2}{I_{rs}\over 2}\psi_{-r}^{\mu}\psi_{-s\mu}+
\sum_{n\ge 1}{(-1)^n\over 2n}(j_{-n})^2
-\sum_{k\ge 1}{(-1)^k\over k}j_{-2k}+\sum_{n\ge
1}(-1)^n\eta_{-n}\xi_{-n},\nonumber
\end{eqnarray}
where the oscillators are given by
\begin{eqnarray}
&&\psi^{\mu}(z)=\sum_r\psi^{\mu}_rz^{-r-1/2},
~~~~~~\{\psi^{\mu}_r,\psi^{\nu}_s\}=\eta^{\mu\nu}\delta_{r+s,0},\\
&&\phi(z)=\hat{\phi}_0-j_0 \log z+\sum_{n\ne 0}{1\over
 n}j_nz^{-n},~~~~~~~
[j_0,\hat{\phi}_0]=1,~~~[j_m,j_n]=-m\delta_{m+n,0},\\
&&\xi(z)=\sum_n\xi_nz^{-n}\,,~~~\eta(z)=\sum_n\eta_nz^{-n-1}\,,
~~~\{\xi_m,\eta_n\}=\delta_{m+n,0}\,,
\end{eqnarray}
and the vacuum with $\phi$-charge $q$ is 
defined by $|q\rangle=e^{q\phi}(0)|0\rangle
=e^{q\hat{\phi}_0}|0\rangle$.

Combining (\ref{eq:IXbc}) and  (\ref{eq:Ipsiphi}), the identity string
field in open superstring field theory in the large Hilbert space is
obtained:
\begin{eqnarray}
\label{eq:Itotal}
 |I\rangle&=&|I\rangle_b\otimes |I\rangle_f.
\end{eqnarray}
The index $\mu$ in the exponent of $|I\rangle_b$ runs over $0,1,\cdots
,9$ and $|p^{\mu}=0\rangle \otimes |q=0\rangle$ is the conformal vacuum.
$|I\rangle$  is Grassmann even and has both ghost and picture number $0$. 
BRS invariance $\Q|I\rangle=0$ follows from the construction of LPP vertex
and $\eta_0|I\rangle=0$ can be checked directly.
We can easily derive the following connection conditions of each oscillators
on the  identity  string field $|I\rangle$ using the above explicit 
expression:\footnote{
For a derivation using CFT, see ref.~\cite{Schnabl:2002gg}.
}
\begin{eqnarray}
\label{eq:conn1}
&&(\alpha_n^{\mu}+(-1)^n\alpha_{-n}^{\mu})|I\rangle=0,~~~~
(b_n-(-1)^nb_{-n})|I\rangle=0,\\
\label{eq:conn2}
&&(c_{2k}+c_{-2k}-(-1)^k2 c_0)|I\rangle=0,~~~~
(c_{2k+1}-c_{-(2k+1)}-(-1)^k(c_1-c_{-1}))|I\rangle=0,~~~~~\\
\label{eq:conn3}
&&\biggl(\psi^{\mu}_r-\sum_{s\ge 1/2}I_{rs}\psi^{\mu}_{-s}\biggr)
|I\rangle=0,~~
(\xi_n-(-1)^n\xi_{-n})|I\rangle=0,~~
(\eta_n+(-1)^n\eta_{-n})|I\rangle=0,~~~~\\
\label{eq:conn4}
&&(j_{2k}+j_{-2k}-(-1)^k 2)|I\rangle=0,~~
(j_{2k-1}-j_{-(2k-1)})|I\rangle=0,~~(k\ge 1);~~~~j_0|I\rangle=0.
\end{eqnarray}
The identity string field $|I\rangle$ satisfies the reality condition:
$(|I\rangle)^{\dagger}={\rm bpz}(|I\rangle)$, where the BPZ conjugation
is given by ${\rm bpz}(\sigma_n)=(-1)^{-n+h}\sigma_{-n}$
for oscillators of a primary field $\sigma(z)$ with conformal dimension
$h$ and ${\rm
bpz}(|p^{\mu};q\rangle)=(|-p^{\mu};q\rangle)^{\dagger}$ for
zero mode part and use has been made of $(-1)^{r+s}I_{rs}=I_{rs}$.
We note that the identity string field $|I\rangle$ can be rewritten
as
\begin{eqnarray}
 |I\rangle&=&{(2i)^{1\over 4}\over 4i}b(i\pi/2)b(-i\pi/2)
:\!e^{{1\over 2}\phi(i\pi/2)}\!::\!e^{{1\over 2}\phi(-i\pi/2)}\!:
e^{E'}\!
c_0c_1|p^{\mu}=0;q=-1\rangle\,,\\
E'&=&-{1\over 2}\sum_{n\ge 1}{(-1)^n\over n}\alpha_{-n}^{\mu}
\alpha_{-n\mu}
+\sum_{n\ge 1}(-1)^nc_{-n}b_{-n}\nonumber\\
&&+{1\over 2}\sum_{r,s\ge 1/2}I_{rs}\psi_{-r}^{\mu}\psi_{-s\mu}
+{1\over 2}\sum_{n\ge 1}{(-1)^n\over n}j_{-n}j_{-n}+
\sum_{n\ge 1}(-1)^n\eta_{-n}\xi_{-n}\,,
\end{eqnarray}
where we denoted
\begin{eqnarray}
&&b(i\sigma)=\sum_{n}b_ne^{-in\sigma},\\
&&:e^{q\phi(i\sigma)}\!:=e^{-{1\over 2}q(q+2)i\sigma}
e^{-q\sum_{n\ge 1}{1\over
 n}j_{-n}e^{in\sigma}}e^{q\hat{\phi}_0}e^{-iq\sigma j_0}
e^{q\sum_{n\ge 1}{1\over n}j_ne^{-in\sigma}}\,.
\label{eq:eqphi}
\end{eqnarray}
The extra factor $e^{-{1\over 2}q(q+2)i\sigma}$ in 
the normal order form comes from the conformal factor 
under the map  $z=e^{\rho}$.

%%%%%%%%%%%%%%%%%%%%%%%%%%%%%%%%%%%%%%%%%%%%%%%%%%%%%%%%%%%%%%%%
\section{Action around a classical solution
\label{sec:aacs}
}

The Berkovits' action for open superstring field theory in the NS
sector is given by
\begin{eqnarray}
\label{eq:WZWaction}
 S[\Phi]&=&-{1\over 2g^2}\langle\!\langle
\bar{A}_{\eta_0}\bar{A}_{Q}
\rangle\!\rangle
-{1\over 2g^2}\int^1_0 dt\,\langle\!\langle
A_t\{A_Q,A_{\eta_0}\}\rangle\!\rangle\,,
\end{eqnarray}
where $A_{\eta_0},A_Q$ and $A_t$ are defined by string field $\Phi(t)$
parametrized by $t$ with boundary value $\Phi(1)=\Phi,\Phi(0)=0$ :
\begin{eqnarray}
&&A_{\eta_0}=e^{-\Phi(t)}(\eta_0 e^{\Phi(t)})\,,~~~~
A_Q=e^{-\Phi(t)}(Q e^{\Phi(t)})\,,~~~~
A_t=e^{-\Phi(t)}(\partial_t e^{\Phi(t)})\,,
\end{eqnarray}
and $\bar{A}_{\eta_0}=A_{\eta_0}|_{t=1},\bar{A}_Q=A_Q|_{t=1}$.
We usually take $\Phi(t)=t\Phi$ although the action $S[\Phi]$
itself does not depend on this parameterization.
We often denote $\{A,B\}=AB+BA$ and $[A,B]=AB-BA$ and 
omit the symbol for the star product among string fields.
We note that $\eta_0,Q$ and $\partial_t$ are derivations with respect
to the star product:
\begin{eqnarray}
 \eta_0 (A*B)&=&(\eta_0A)*B+(-1)^{|A|}A*(\eta_0 B),\\
Q(A*B)&=&(Q A)*B+(-1)^{|A|}A*(Q B)\,,\\
\partial_t(A*B)&=&(\partial_t A)*B+A*(\partial_t B)\,,
\end{eqnarray}
where $(-1)^{|A|}$ is $+1(-1)$ when $A$ is Grassmann even (odd)
and have nilpotency $\eta_0^2=Q^2=0$ and (anti-)commutativity:
$\{\eta_0,Q\}=0,[\partial_t,\eta_0]=[\partial_t,Q]=0$.
The above WZW type action (\ref{eq:WZWaction}) can be rewritten in a
rather simple form \cite{rf:BOZ}:
\begin{eqnarray}
\label{eq:BOZ}
 S[\Phi]&=&-{1\over g^2}\int_0^1dt\langle\!\langle (\eta_0
  A_t)A_Q\rangle\!\rangle\,.
\end{eqnarray}
Let us consider re-expansion of this action 
 around $\Phi^{(0)}$ with respect to $\Phi'$ in the sense
 $e^{\Phi(t)}=e^{\Phi^{(0)}(t)}e^{\Phi'(t)}$.
The integrand of (\ref{eq:BOZ}) can be rewritten as:
\begin{eqnarray}
\langle\!\langle(\eta_0 A_t)A_Q\rangle\!\rangle
&=&\langle\!\langle(e^{-\Phi'(t)}(\eta_0 A_t^{(0)})e^{\Phi'(t)}
+\eta_0A_t'+(\eta_0 e^{-\Phi'(t)})A_t^{(0)}e^{\Phi'(t)}
+e^{-\Phi'(t)}A_t^{(0)}(\eta_0 e^{\Phi'(t)}))\nonumber\\
&&\times (e^{-\Phi'(t)}A_Q^{(0)}e^{\Phi'(t)}+A'_Q)\rangle\!\rangle
\nonumber\\
&=&\langle\!\langle(\eta_0 A_t^{(0)})A_Q^{(0)}\rangle\!\rangle+
\langle\!\langle(\eta_0 A_t^{\prime})A_Q^{\prime}\rangle\!\rangle
\nonumber\\
&&+\langle\!\langle (\eta_0 A'_t)e^{-\Phi'(t)}A_Q^{(0)}e^{\Phi'(t)}
-e^{\Phi'(t)}(\eta_0 e^{-\Phi'(t)})\partial_t A_Q^{(0)}\rangle\!\rangle
\nonumber\\
&&+\langle\!\langle(\eta_0A_t^{(0)})e^{\Phi'(t)}A'_Qe^{-\Phi'(t)}
+(\eta_0 e^{-\Phi'(t)})A_t^{(0)}e^{\Phi'(t)}A'_Q
+e^{-\Phi'(t)}A_t^{(0)}(\eta_0e^{\Phi'(t)})A'_Q\rangle\!\rangle
\nonumber\\
&&+\langle\!\langle e^{-\Phi'(t)}(\eta_0e^{\Phi'(t)})(QA_t^{(0)})
\rangle\!\rangle\,,
\label{eq:integrand}
\end{eqnarray}
where we denote $A_t^{(0)}=e^{-\Phi^{(0)}(t)}(\partial_t
e^{\Phi^{(0)}(t)})$, $A_Q^{(0)}=e^{-\Phi^(0)(t)}(Q e^{\Phi^{(0)}(t)})$,
$A'_t=e^{-\Phi'(t)}(\partial_t e^{\Phi'(t)})$, and $A'_Q=e^{-\Phi'(t)}(Q
e^{\Phi'(t)})$
and use has been made of cyclic property:
\begin{eqnarray}
&&\langle\!\langle A_1\cdots A_{n-1}\Phi\rangle\!\rangle
=\langle\!\langle\Phi A_1\cdots A_{n-1}\rangle\!\rangle\,,\\
&&\langle\!\langle A_1\cdots A_{n-1}(Q\Phi)\rangle\!\rangle
=-\langle\!\langle (Q\Phi) A_1\cdots A_{n-1}\rangle\!\rangle\,,\\
&&\langle\!\langle A_1\cdots A_{n-1}(\eta_0\Phi)\rangle\!\rangle
=-\langle\!\langle (\eta_0\Phi) A_1\cdots A_{n-1}\rangle\!\rangle\,,
\end{eqnarray}
and an identity $[A_t^{(0)},A_Q^{(0)}]=QA_t^{(0)}-\partial_t A_Q^{(0)}$.
Using $e^{\Phi'(t)}(\eta_0A_t')e^{-\Phi'(t)}=-\partial_t(e^{\Phi'(t)}(\eta_0
e^{-\Phi'(t)}))$, $Q(e^{\Phi'(t)}(\eta_0 e^{-\Phi'(t)}))
=e^{\Phi'(t)}(\eta_0A_Q')e^{-\Phi'(t)}$ and partial integrability:
\begin{eqnarray}
 &&\langle\!\langle Q(\cdots)\rangle\!\rangle=0\,,~~~~
\langle\!\langle \eta_0(\cdots)\rangle\!\rangle=0\,,
\end{eqnarray}
we can simplify
 the last three lines of (\ref{eq:integrand}) as:
\begin{eqnarray}
&&\langle\!\langle (\eta_0 A'_t)e^{-\Phi'(t)}A_Q^{(0)}e^{\Phi'(t)}
-e^{\Phi'(t)}(\eta_0 e^{-\Phi'(t)})\partial_t A_Q^{(0)}\rangle\!\rangle
=-\partial_t\langle\!\langle e^{\Phi'(t)}(\eta_0e^{-\Phi'(t)})
A_Q^{(0)}\rangle\!\rangle,\\
&&\langle\!\langle(\eta_0A_t^{(0)})e^{\Phi'(t)}A'_Qe^{-\Phi'(t)}
+(\eta_0 e^{-\Phi'(t)})A_t^{(0)}e^{\Phi'(t)}A'_Q
+e^{-\Phi'(t)}A_t^{(0)}(\eta_0e^{\Phi'(t)})A'_Q\rangle\!\rangle\nonumber\\
&&~~~~=-\langle\!\langle A_t^{(0)}e^{\Phi'(t)}(\eta_0A_Q')e^{-\Phi'(t)}
\rangle\!\rangle\,,\\
&&\langle\!\langle e^{-\Phi'(t)}(\eta_0e^{\Phi'(t)})(QA_t^{(0)})
\rangle\!\rangle=\langle\!\langle
A_t^{(0)}e^{\Phi'(t)}(\eta_0A_Q')e^{-\Phi'(t)}
\rangle\!\rangle\,.
\end{eqnarray}
Then we have proved an identity
\begin{eqnarray}
 \langle\!\langle(\eta_0 A_t)A_Q\rangle\!\rangle
=\langle\!\langle(\eta_0 A_t^{(0)})A_Q^{(0)}\rangle\!\rangle+
\langle\!\langle(\eta_0 A_t^{\prime})A_Q^{\prime}\rangle\!\rangle
-\partial_t\langle\!\langle e^{\Phi'(t)}(\eta_0e^{-\Phi'(t)})
A_Q^{(0)}\rangle\!\rangle\,,
\end{eqnarray}
which implies the action (\ref{eq:BOZ}) is rewritten for 
$e^{\Phi}=e^{\Phi^{(0)}}e^{\Phi'}$ as
\begin{eqnarray}
 S[\Phi]=S[\Phi^{(0)}]+S[\Phi']+{1\over g^2}\langle\!\langle
e^{\Phi'}(\eta_0e^{-\Phi'})\bar{A}_Q^{(0)}\rangle\!\rangle\,,
\label{eq:action12}
\end{eqnarray}
where we have imposed ordinary boundary conditions
 $\Phi^{(0)}(1)=\Phi^{(0)},\Phi'(1)=\Phi',\Phi^{(0)}(0)=\Phi'(0)=0$
and denoted $\bar{A}_Q^{(0)}=A_Q^{(0)}|_{t=1}$.
The last extra term of the above action can be rewritten as follows:
\begin{eqnarray}
&&{1\over g^2}\langle\!\langle
e^{\Phi'}(\eta_0e^{-\Phi'})\bar{A}_Q^{(0)}\rangle\!\rangle\nonumber\\
&=&{1\over g^2}\int_0^1dt\partial_t
\langle\!\langle
e^{\Phi'(t)}(\eta_0e^{-\Phi'(t)})\bar{A}_Q^{(0)}\rangle\!\rangle
=-{1\over g^2}\int_0^1dt\langle\!\langle e^{\Phi'(t)}(\eta_0A_t')
e^{-\Phi'(t)}\bar{A}_Q^{(0)}\rangle\!\rangle\nonumber\\
&=&-{1\over g^2}\int_0^1dt\langle\!\langle (\eta_0A_t')
(e^{-\Phi'(t)}\bar{A}_Q^{(0)}e^{\Phi'(t)}-\bar{A}_Q^{(0)})
-A_t'(\eta_0\bar{A}_Q^{(0)})\rangle\!\rangle\,.
\label{eq:otsuri}
\end{eqnarray}
In the first equality, we have kept $\bar{A}_Q^{(0)}$ intact
in $t$-integration and used
$e^{\Phi'(t)}(\eta_0A_t')e^{-\Phi'(t)}=-\partial_t(e^{\Phi'(t)}(\eta_0
e^{-\Phi'(t)}))$ again in the second equality. Using
(\ref{eq:BOZ}), (\ref{eq:action12}) and (\ref{eq:otsuri}),
and imposing equation of motion for $\Phi^{(0)}$:
$\eta_0\bar{A}_Q^{(0)}=0$, we have obtained the action for $\Phi'$
around a classical solution $\Phi^{(0)}$ in the same form as the original one:
\begin{eqnarray}
S'[\Phi']&\equiv&S[\Phi]-S[\Phi^{(0)}]=-{1\over
 g^2}\int_0^1dt\langle\!\langle (\eta_0
 A'_t)A'_{Q'}\rangle\!\rangle\,,
\end{eqnarray}
with $A'_{Q'}=e^{-\Phi'(t)}(Q'e^{\Phi'(t)})$, where
the new BRS operator $Q'$ is given by
\begin{eqnarray}
\label{eq:Q_B'formula}
 Q'B&=&QB+\bar{A}_Q^{(0)}*B-(-1)^{|B|}B*\bar{A}_Q^{(0)},
~~~~~~~\bar{A}_Q^{(0)}=e^{-\Phi^{(0)}}(Qe^{\Phi^{(0)}})\,.
\end{eqnarray}
We note that $Q'$ is a derivation with respect to the star product,
nilpotency $Q^{\prime 2}=0$ holds automatically,
and $\{Q',\eta_0\}=0$ is satisfied by 
equation of motion $\eta_0\bar{A}_Q^{(0)}=0$ for $\Phi^{(0)}$.
The above action can be rewritten in the ordinary WZW form again
\begin{eqnarray}
\label{eq:WZWaction_p}
 S'[\Phi']&=&-{1\over 2g^2}\langle\!\langle
\bar{A}'_{\eta_0}\bar{A}'_{Q'}
\rangle\!\rangle
-{1\over 2g^2}\int^1_0 dt\,\langle\!\langle
A'_t\{A'_{Q'},A'_{\eta_0}\}\rangle\!\rangle\,,
\end{eqnarray}
where  $A'_{\eta_0}=e^{-\Phi'(t)}(\eta_0e^{\Phi'(t)}),\,
\bar{A}'_{\eta_0}=A'_{\eta_0}|_{t=0},\,\bar{A}'_{Q'}=A'_{Q'}|_{t=1}$,
using the method in \cite{rf:BOZ}.

%%%%%%%%%%%%%%%%%%%%%%%%%%%%%%%%%%%%%%%%%%%%%%%%%%%%%%%%%%%%%%%%
\section{Supersymmetry in superstring field theories}

First, we will show that the fermionic transformation
(\ref{Eq:susytrans}), which is generated by (\ref{eq:Oep}), corresponds
to global space-time supersymmetries.  

For the parameter (\ref{eq:Oep}), the transformation law 
given in (\ref{Eq:fermionicsym}) and (\ref{eq:fermionic_g2}) becomes
\begin{eqnarray}
 \delta_{\epsilon}\Phi&=&-{{\rm ad}_{\Phi}\over 1-e^{-{\rm ad}_{\Phi}}}
\{\Omega(\epsilon),\eta_0\Psi\}
=-{{\rm ad}_{\Phi}\over 1-e^{-{\rm ad}_{\Phi}}}
\left(\oint{dz\over 2\pi i}
\epsilon_{\alpha}\xi S^{\alpha}_{(-1/2)}(z)\eta_0\Psi\right)
\,,\\
\delta_{\epsilon}(\eta_0\Psi)&=&\eta_0\{\Omega(\epsilon),e^{-\Phi}\Q
 e^{\Phi}\}
=
\eta_0\oint{dz\over 2\pi i}\epsilon_{\alpha}\xi
S^{\alpha}_{(-1/2)}(z)(e^{-\Phi}\Q e^{\Phi})\,,
\end{eqnarray}
where the term $\eta_0\Q \Omega(\epsilon)$ 
is not included as is explained in
footnote \ref{footnote:susy} in order to express
it only in terms of a contour integration (\ref{eq:Sepsilon}).
These transformations are equivalent to (\ref{Eq:susytrans}).
The equations of motion are given by
\begin{eqnarray}
\label{eq:EOMf1}
f_1&\equiv&\eta_0(e^{-\Phi}\Q e^{\Phi})+(\eta_0\Psi)^2=0,\\
\label{eq:EOMf2}
f_2&\equiv&e^{-\Phi}(\Q(e^{\Phi}(\eta_0\Psi)e^{-\Phi}))e^{\Phi}
=\Q\eta_0\Psi+\{e^{-\Phi}\Q e^{\Phi},\eta_0\Psi\}=0\,.
\end{eqnarray}
Note that the R sector string field $\Psi$ is 
involved in the equations of motion through the particular form
$\eta_0\Psi$.
We apply the transformation (\ref{Eq:susytrans}) to the string fields
$f_1$ and $f_2$:
\begin{eqnarray}
 \delta_{\epsilon}f_1&=&\eta_0{\cal S}(\epsilon)f_2\,,\\
\delta_{\epsilon}f_2&=&-\{\Q,{\cal S}(\epsilon)\}f_1
+[f_1,{\cal S}(\epsilon)(e^{-\Phi} \Q e^{\Phi})]
+\{{\cal S}(\epsilon)f_2,\,\eta_0\Psi\}.
\end{eqnarray}
If $f_1=f_2=0$, we find $\delta_\epsilon f_1=\delta_\epsilon
f_2=0$. Hence this symmetry is realized only on-shell. 

Let us consider how massless fields are transformed by
(\ref{Eq:susytrans}). The string fields contain massless fields as
follows:
\begin{eqnarray}
 |\Phi_A\rangle
&=&\int{d^{10}p\over
  (2\pi)^{10}}(\tilde{A}_{\mu}(p)c\xi e^{-\phi}\psi^{\mu}(0)
+\tilde{B}(p)c\partial c \xi\partial \xi e^{-2\phi}(0))
|p^{\mu},q=0\rangle,\\
|\Psi_{\lambda}\rangle&=&\int{d^{10}p\over
  (2\pi)^{10}}\tilde{\lambda}_{\alpha}(p)\xi S^{\alpha}_{(-1/2)}c(0)
|p^{\mu},q=0\rangle
\,,
\end{eqnarray}
where $\tilde{A}_{\mu}(p),\tilde{B}(p)$ and
$\tilde{\lambda}_{\alpha}(p)$ 
denote Fourier transforms of gluon,
auxiliary Nakanishi-Lautrup \cite{rf:BS} and gluino \cite{rf:Rsector} fields,
respectively. We note that $q$ in $|p^{\mu},q\rangle$
implies a zero-mode momentum of $\phi$.
Applying the transformation (\ref{Eq:susytrans}) to these 
fields, we find
\begin{eqnarray}
\label{Eq:delNS}
\delta_{\epsilon}|\Phi_A\rangle&=&\oint{dz\over 2\pi i}
\epsilon_{\alpha}\xi S^{\alpha}_{(-1/2)}(z)\eta_0|\Psi_{\lambda}\rangle
 +\cdots \nonumber\\
&=&\int{d^{10}p\over
  (2\pi)^{10}}(-i\epsilon_{\alpha}(\Gamma_{\mu}C)^{\alpha\beta}
\tilde{\lambda}_{\beta}(p))c\xi 
e^{-\phi}\psi^{\mu}(0)|p^{\mu},0\rangle
+\cdots \,,\\
\delta_{\epsilon}(\eta_0|\Psi_{\lambda}\rangle)&=&
\eta_0\oint{dz\over 2\pi i}
\epsilon_{\alpha}\xi S^{\alpha}_{(-1/2)}(z)\Q|\Phi_A\rangle
 +\cdots \\
\label{Eq:delR}
&=&\int{d^{10}p\over
  (2\pi)^{10}}\epsilon_{\alpha}
\Biggl({1\over 4}\sqrt{2\alpha'}(p_{\mu}\tilde{A}_{\nu}(p)
-p_{\nu}\tilde{A}_{\mu}(p))(\Gamma^{\mu\nu})^{\alpha}_{~\beta}
\nonumber\\
&&+\Biggl(
\sqrt{\alpha'\over 2}p^{\mu}\tilde{A}_{\mu}(p)
+\tilde{B}(p)\Biggr)
\delta^{\alpha}_{\beta}
\Biggr)S^{\beta}_{(-1/2)}c(0)|p^{\mu},0\rangle
+\cdots,
\end{eqnarray}
where $(+\cdots )$ denotes quadratic or higher order terms 
with respect to component fields.
Here we have calculated the above results by using the OPEs,\footnote{
We have used the convention in \cite{W} for spin fields
and taken $T^{\rm m}(z)=-{1\over 4\alpha'}
\partial X_{\mu}\partial X^{\mu}(z)-{1\over
2}\psi^{\mu}\partial\psi_{\mu}(z),~G^{\rm m}(z)
={i\over \sqrt{2\alpha'}}\psi^{\mu}\partial X_{\mu}(z)$
with
$X^{\mu}(y)X^{\nu}(z)\sim -2\alpha'\eta^{\mu\nu}\log(y-z),~
\psi^{\mu}(y)\psi^{\nu}(z)\sim \eta^{\mu\nu}(y-z)^{-1}$ 
in the matter sector
and $j_{\rm B}(z)=
c\left(T^{\rm m}-{1\over 2}(\partial\phi)^2
-\partial^2\phi+\partial \xi\eta
\right)\!(z)+bc\partial c(z)+\eta e^{\phi}G^{\rm m}(z)-\eta\partial\eta
e^{2\phi}b(z)+\partial^2c(z)+\partial(c\xi\eta)(z)
$ for the BRS current.
}
\begin{eqnarray}
 &&S^{\alpha}_{(-1/2)}(y)S^{\beta}_{(-1/2)}(z)\sim
{1\over y-z}i(\Gamma_{\mu}C)^{\alpha\beta}\psi^{\mu}e^{-\phi}(z)\,,\\
&&j_{\rm B}(y)c\xi e^{-\phi}\psi^{\mu}e^{ip_{\nu} X^{\nu}}(z)\sim
{1\over (y-z)^2}\sqrt{2\alpha'}p^{\mu}c e^{ip_{\nu}X^{\nu}}(z)
+{1\over y-z}\biggl(
{i\over \sqrt{2\alpha'}}c\partial X^{\mu}+\eta
c e^{\phi}\psi^{\mu}\nonumber\\
&&~~~~~~~~~~~~~~~~~~~~~~~
+\sqrt{2\alpha'}c(p_{\nu}\psi^{\nu}\psi^{\mu}
+p^{\mu}(\partial\phi-\xi\eta))
-\alpha'p^2c\partial c\xi e^{-\phi}\psi^{\mu}
\biggr) e^{ip_{\nu} X^{\nu}}(z),\\
&&j_{\rm B}(y)c\partial c \xi\partial \xi e^{-2\phi}
e^{ip_{\nu} X^{\nu}}(z)\sim {1\over (y-z)^2}ce^{ip_{\nu} X^{\nu}}(z)\\
&&~~~~~~~~~~~~~~~~~~~~~~~~~~~~~~
+{1\over y-z}(-\partial c+2c(\partial \phi-\xi\eta)
+\sqrt{2\alpha'}c\partial c \xi p_{\mu}\psi^{\mu}e^{-\phi})
e^{ip_{\nu} X^{\nu}}(z),
\nonumber\\
&&\xi S^{\alpha}_{(-1/2)}(y)c(\partial \phi-\xi\eta)(z)\sim {1\over y-z}
\left({1\over 2}\xi S^{\alpha}_{(-1/2)}c\right)(z)\,,\\
&&\xi S_{(-1/2)}^{\alpha}(y)c\psi^{\nu}\psi^{\mu}(z)\sim {1\over y-z}
{1\over 2}\xi(\Gamma^{\nu\mu}S_{(-1/2)})^{\alpha}c(z)\,.
\end{eqnarray}
From (\ref{Eq:delNS}) and (\ref{Eq:delR}), we can read off the
transformation law for massless fields:
\begin{eqnarray}
\label{Eq:delA}
\delta_{\epsilon}\tilde{A}_{\mu}(p)&=&-i\epsilon
\Gamma_{\mu}C\tilde{\lambda}(p)+\cdots,\\
\label{Eq:delB}
\delta_{\epsilon}\tilde{B}(p)&=&0+\cdots,\\
\label{Eq:dellambda}
\delta_{\epsilon}\tilde{\lambda}_{\alpha}(p)&=&
-{\sqrt{2\alpha'}\over 4}(p_{\mu}\tilde{A}_{\nu}(p)
-p_{\nu}\tilde{A}_{\mu}(p))(\epsilon\Gamma^{\mu\nu})_{\alpha}
-\Biggl(
\sqrt{\alpha'\over 2}p^{\mu}\tilde{A}_{\mu}(p)
+\tilde{B}(p)\Biggr)\epsilon_{\alpha}+\cdots.~~~~
\end{eqnarray}

For massless fields, the linearized equations of motion are calculated as
\begin{eqnarray}
 \Q \eta_0|\Phi_A\rangle&=&\int\!{d^{10}p\over
  (2\pi)^{10}}\biggl(-(\sqrt{2\alpha'}p^{\mu}\tilde{A}_{\mu}(p)+2\tilde{B}(p)
)c\eta(0)\nonumber\\
&&
+(\alpha'p^2\tilde{A}(p)+\sqrt{2\alpha'}p_{\mu}\tilde{B}(p))c\partial c
e^{-\phi}\psi^{\mu}(0)\biggr)|p^{\nu},0\rangle=0,\\
\Q\eta_0|\Psi_{\lambda}\rangle&=&\int\!{d^{10}p\over
  (2\pi)^{10}}
(\alpha'p^2\tilde{\lambda}_{\alpha}(p)
S^{\alpha}_{(-1/2)}c\partial c(0)\nonumber\\
&&
+\sqrt{\alpha'}ip_{\mu}\tilde{\lambda}_{\alpha}(p)
(\Gamma^{\mu})^{\alpha}_{~\dot{\beta}}S^{\dot{\beta}}_{(1/2)}\eta c(0)
)|p^{\nu},0\rangle=0,~~~
\end{eqnarray}
where $S^{\dot{\beta}}_{(1/2)}$ is the  GSO$(+)$ spin operator
with $\phi$-charge $1/2$, dimension $0$ and negative chirality.
Consequently, we obtain the linearized equations of motion:
\begin{eqnarray}
&&\tilde{B}(p)=-\sqrt{\alpha'\over 2}p^{\mu}\tilde{A}_{\mu}(p),~~~~
(p^2\delta_{\mu}^{\nu}-p_{\mu}p^{\nu})\tilde{A}_{\nu}(p)=0,\\
&&\tilde{\lambda}_{\alpha}(p)\,p^2=0\,,~~~~~
\tilde{\lambda}(p)\Gamma^{\mu}p_{\mu}=0\,.
\end{eqnarray}
Under these on-shell conditions, the transformation laws (\ref{Eq:delA}),
(\ref{Eq:delB}) and (\ref{Eq:dellambda}) become
\begin{eqnarray}
 \delta_{\epsilon}A_{\mu}=-i\epsilon \Gamma_{\mu}C\lambda\,,~~~
\delta_{\epsilon}\lambda={i\over 2}\sqrt{\alpha'\over 2}F_{\mu\nu}(\epsilon
\Gamma^{\mu\nu})\,,~~~~
(F_{\mu\nu}=\partial_{\mu}A_{\nu}-\partial_{\nu}A_{\mu}).
\end{eqnarray}
These are nothing but supersymmetry transformation
of 10D supersymmetric Maxwell theory. Hence, the transformation law
(\ref{Eq:susytrans}) contains space-time supersymmetries.

Finally, we would like to comment on supersymmetry in the
cubic open 
superstring field theory \cite{rf:SCSFT} and the modified cubic theory
\cite{rf:PTY,rf:Arefeva}.
\paragraph{Cubic version}
In the Witten's  open superstring field theory, there are
fermionic gauge symmetry and global supersymmetry
at least formally \cite{rf:SCSFT}.
Fermionic gauge symmetry, which is generated by
Grassmann even gauge parameter $\chi$ in the
Ramond sector with picture number $-1/2$ and ghost number $0$, is given by
\begin{eqnarray}
\delta_{\chi} A&=&\Psi*\chi-\chi*\Psi\,,\\
\delta_{\chi} \Psi&=&\Q \chi+X(i)(A*\chi-\chi*A)\,,
\end{eqnarray}
where $X(i)$ is the picture changing operator at the midpoint,
and $A\ (\Psi)$ denotes a Grassmann odd string field in the NS (R)
sector with picture 
number $-1\ (-1/2)$ and ghost number $1\ (1)$.
Formally, by taking 
\begin{eqnarray}
 \chi&=&-\int_{C_{\rm left}}{dz\over 2\pi
  i}\epsilon_{\alpha}S^{\alpha}_{(-1/2)}(z)I,
\end{eqnarray}
and omitting $\Q \chi={1\over 2\pi i}
\epsilon_{\alpha}(cS^{\alpha}_{(-1/2)}(i)-cS^{\alpha}_{(-1/2)}(-i))I$
in $\delta_{\chi}\Psi$, which is itself a symmetry of the action,
the above gauge
transformation becomes global space-time
supersymmetry transformation:
\begin{eqnarray}
 \delta_{\epsilon}A&=&\oint {dz\over 2\pi
  i}\epsilon_{\alpha}S^{\alpha}_{(-1/2)}(z)\Psi\,,\\
\delta_{\epsilon}\Psi&=&X(i)\oint {dz\over 2\pi
  i}\epsilon_{\alpha}S^{\alpha}_{(-1/2)}(z)A\,.
\end{eqnarray}

\paragraph{Modified cubic version}
In modified version of cubic open superstring field theory, there
are also fermionic gauge symmetry and global supersymmetry.
Fermionic gauge symmetry, which is generated by
Grassmann even gauge parameter $\chi$ in the
Ramond sector with picture number $-1/2$ and ghost number $0$, is given by
\begin{eqnarray}
 \delta_{\chi} A&=&X(i)(\Psi*\chi-\chi*\Psi)\,,\\
\delta_{\chi} \Psi&=&\Q\chi+A*\chi-\chi*A\,,
\end{eqnarray}
where $A\ (\Psi)$ denotes a Grassmann odd string field in the NS (R)
sector with picture number $0\ (-1/2)$ and ghost number $1\ (1)$.
Formally, by taking 
\begin{eqnarray}
 \chi&=&-Y(i)\int_{C_{\rm left}}{dz\over 2\pi
  i}\Q \epsilon_{\alpha}\xi S^{\alpha}_{(-1/2)}(z)I=Y(i)
\int_{C_{\rm left}}{dz\over 2\pi i}\epsilon_{\alpha}W^{\alpha}(z)I,\\
&&W^{\alpha}(z)\equiv[\Q,\xi S^{\alpha}_{(-1/2)}(z)]\,,
\end{eqnarray}
with the inverse picture changing operator $Y(z)$,
which is global: $\Q \chi=0$, the above gauge transformation yields
the global space-time supersymmetry transformation
\cite{KO_D}:\footnote{
Similar formula can be found in \cite{rf:Arefeva,Arefeva:1989cm}.
See also \cite{Urosevic:1990as}.
}
\begin{eqnarray}
 \delta_{\epsilon}A&=&\oint{dz\over 2\pi
  i}\epsilon_{\alpha}W^{\alpha}(z)\Psi,\\
 \delta_{\epsilon}\Psi&=&Y(i)\oint{dz\over 2\pi
  i}\epsilon_{\alpha}W^{\alpha}(z)A.
\end{eqnarray}

%%%%%%%%%%%%%%%%%%%%%%%%%%%%%%%%%%%%%%%%%%%%%%%%%%%%%%%%%%%%%%%%%%%%%%%%%%%%%
\section{Classical solutions and marginal deformations
 in Witten's bosonic open string field theory\label{sec:bosonic}}

In this section, we consider classical solutions
of the Witten's bosonic string field theory
corresponding to marginal deformations,
which are generalization of the previous ones investigated in
\cite{rf:TT1}\footnote{
General arguments in bosonic string field theory are given in
\cite{Kluson_M}.
} and counterparts of the arguments in \S \ref{sec:super_gen}.
 We discuss a class of solutions using 
a current $J^a$ associated with a Lie algebra ${\cal G}$.
We suppose the OPE among currents with adjoint indices of the
form
\begin{eqnarray}
\label{eq:OPE_JJb}
 J^a(y)J^b(z)&\sim&-g^{ab}{1\over (y-z)^2}+{1\over
  y-z}f^{ab}_{~~c}J^c(z)\,,\\
g^{ab}&=&{1\over 2}(f^{ac}_{~~d}f^{bd}_{~~c}-\Omega^{ab})\,,
\end{eqnarray}
where $f^{ab}_{~~c}$ is the structure constant of ${\cal G}$ and 
 $\Omega^{ab}$ is a particular invertible invariant matrix such as
 (\ref{eq:Oinvariance}).
$J^a$ is a primary field with dimension 1 for the energy momentum
tensor given by the Sugawara form:
\begin{eqnarray}
\label{eq:Sugawara_b}
 T(z)&=&\Omega_{ab}:J^aJ^b:(z),~~~~~(\,\Omega^{ab}\Omega_{bc}=\delta^a_c\,)\,.
\end{eqnarray}
In fact, we can show the OPEs
\begin{eqnarray}
T(y)J^a(z)&\sim&{1\over (y-z)^2}J^a(z)+{1\over y-z}\partial J^a(z),\\
T(y)T(z)&\sim&{c\over
 2}{1\over (y-z)^4}+{1\over (y-z)^2} 2T(z)+{1\over y-z}\partial T(z),
\end{eqnarray}
where the central charge $c$ of the Virasoro algebra is given by
$c={\rm dim}{\cal G}-f^{ac}_{~~d}f^{bd}_{~~c}\Omega_{ab}$ \cite{Moh}.
In the following, we assume that the background is described by 
the above CFT with $c=26$, and construct a classical
solution of bosonic open string field theory of cubic form:
\begin{eqnarray}
\label{eq:action_b}
 S[\Psi]&=&-{1\over g^2}\left({1\over 2}\langle \Psi \Q\Psi\rangle 
+{1\over 3}\langle \Psi \Psi\Psi\rangle\right)
\end{eqnarray}
on this background. Namely, the BRS operator in the kinetic
term is
\begin{eqnarray}
\label{eq:QB_b}
 \Q&=&\oint{dz\over 2\pi i}\left(cT(z)+bc\partial c(z)\right)
\end{eqnarray}
where $T(z)$ is given by eq.~(\ref{eq:Sugawara_b}) and interaction term
is defined by the Witten $*$ product using conformal mappings and CFT
correlators \cite{LPP1, Sen1989}.
With the above setup, we can show commutation relations
\begin{eqnarray}
&&\{\Q,cJ^a(z)\}=0,~~~~~\{\Q,c(z)\}=c\partial c(z),\\
&&\{V_L^a(f),V_L^b(g)\}={1\over 2}g^{ab}\{\Q,C_L(fg)\},~~~
\{V_L^a(f),C_L(g)\}=\{C_L(f),C_L(g)\}=0,~~
\end{eqnarray}
where $V_L^a(f)=\int_{C_{\rm left}}{dz\over 2\pi i}{1\over \sqrt{2}}f(z)
cJ^a(z)$ and $C_L(f)=\int_{C_{\rm left}}{dz\over 2\pi i}f(z)c(z)$ using
similar method in \cite{rf:TT2}. Then, noting
$cJ^a(z)$ is a primary field with dimension $0$, we have
\begin{eqnarray}
&&V_L^a(F_a)I*V_L^b(F_b)I=
V_L^a(F_a)V_L^b(F_b)I={1\over 4}g^{ab}\{\Q,C_L(F_aF_b)\}I\,,
\end{eqnarray}
with $F_a(-1/z)=z^2F_a(z)$. Using this relation, we can show that 
\begin{eqnarray}
\label{eq:solution_b}
\Psi_0&=&-V^a_L(F_a)I-{1\over 4}g^{ab}C_L(F_aF_b)I\,,
~~~F_a(-1/z)=z^2F_a(z)\,,
\end{eqnarray}
satisfies the equation of motion: $\Q\Psi_0+\Psi_0*\Psi_0=0$.
If we re-expand the action around this solution such as 
$\Psi=\Psi_0+\Psi'$, we have $S[\Psi]=S[\Psi_0]+S'[\Psi']$ where 
the new action $S'[\Psi']$ is the same form as original one
(\ref{eq:action_b}) except that the new BRS operator is given by
$\Q'A=\Q A+\Psi_0*A-(-1)^{|A|}A*\Psi_0$, or more explicitly:
\begin{eqnarray}
\label{eq:QBp_b}
 \Q'&=&\Q-V^a(F_a)-{1\over 4}g^{ab}C(F_aF_b)\,,
\end{eqnarray}
($V^a(f)=\oint{dz\over 2\pi i}{1\over \sqrt{2}}f(z)
cJ^a(z),\,C(f)=\oint{dz\over 2\pi i}f(z)c(z)$).
Comparing this $\Q'$ with original one $\Q$ (\ref{eq:QB_b}), we find
the Virasoro operator $T(z)$ in the matter sector is replaced by 
$T(z)-{1\over\sqrt{2}}F_a(z)J^a(z)-{1\over 4}g^{ab}F_a(z)F_b(z)$.
In fact, if we define $T'(z)=\sum_nL'_nz^{-n-2}$ as
\begin{eqnarray}
\label{eq:Lp_b}
L'_n&=&L_n-{1\over \sqrt{2}}\sum_{k}F_{a,k}J^a_{n-k}-{1\over 4}g^{ab}
\sum_{k}F_{a,n-k}F_{b,k}\,,
\end{eqnarray}
with $F_{a,n}=\oint {d\sigma\over
2\pi}e^{i(n+1)\sigma}F_a(e^{i\sigma})$, we obtain the Virasoro algebra
with the same central charge $c=26$ as the original one:
\begin{eqnarray}
&&[L_m',L_n']=(m-n)L_{m+n}'+{c\over 12}(m^3-m)\delta_{m+n,0}\,.
\end{eqnarray}
Furthermore, by taking $J^{\prime a}(z)=\sum_nJ^{\prime a}_nz^{-n-1}$
as\footnote{In the definition of $M^a_{~b,n}$, we use 
path-ordered form in the same way as eq.~(\ref{eq:path_o}).
Therefore, we have similar formulae to eqs.~(\ref{eq:formula_M1}),
(\ref{eq:formula_M2}) and (\ref{eq:formula_M3})
by replacing $\Omega^{ab}$ with $-2g^{ab}$ in the bosonic case.
Note that $f^{ab}_{~~c}g^{cd}+f^{ad}_{~~c}g^{cb}=0$.}
\begin{eqnarray}
&&J^{\prime a}_n=\sum_kM^a_{~b,k}\left(J^b_{n-k}+{1\over\sqrt{2}}
g^{bc}F_{c,n-k}\right)\,,
\label{eq:mapj_b}\\
&&\sum_nM^a_{~b,n}e^{-in\sigma}=\left[
{\mathbf{P}}\exp\left(i\int_0^1dt\, \sigma
A(t\sigma)\right)\right]^a_{~b}\,,~~~
A^a_{~b}(\sigma)={1\over
\sqrt{2}}f^{ac}_{~~b}e^{i\sigma}F_{c}(e^{i\sigma})\,,
\end{eqnarray}
the same commutation relations as the original one are recovered:
\begin{eqnarray}
&&[J^{\prime a}_m,J^{\prime b}_n]=
-g^{ab}m\delta_{m+n,0}+f^{ab}_{~~c}J^{\prime c}_{m+n},
~~~~[L_m',J^{\prime a}_n]=-nJ^{\prime a}_{m+n}.
\end{eqnarray}
In the above, we have constructed a classical solution $\Psi_0$
(\ref{eq:solution_b}) and re-expanded around it.
The obtained action $S'[\Psi']$ is also reproduced by 
replacing $T(z)$ with $T'(z)$ (\ref{eq:Lp_b}) in the original action
$S[\Psi]$ (\ref{eq:action_b}). This replacement is induced by 
the map $J^a\rightarrow J^{\prime a}$ (\ref{eq:mapj_b})
in terms of the current which preserves the algebra among $(J^a(z),T(z))$.
At least formally, we can show that the vacuum energy
vanishes at this solution \cite{rf:TT2, rf:KugoZwie}:
\begin{eqnarray}
 S[\Psi_0]=\int_0^1 dt{d\over dt}S[\Psi_0(t)]=
-{1\over g^2}\int_0^1 dt\langle
{d\over dt}\Psi_0(t)(\Q\Psi_0(t)+\Psi_0(t)*\Psi_0(t))\rangle=0
\label{eq:vace_b}
\end{eqnarray}
where $\Psi_0(t)$ is given by replacing $F_a(z)$ with $tF_a(z)$ 
in $\Psi_0$ (\ref{eq:solution_b}).
These facts suggest that $\Psi_0$ (\ref{eq:solution_b}) 
may be a pure gauge solution. 
In fact, noting $[\Q,J^a(z)]=\partial(c J^a)(z)$,
we make an ansatz for the gauge parameter as
\begin{eqnarray}
&&\Lambda_LI=\int_{C_{\rm left}}{dz\over 2\pi i}
  {1\over \sqrt{2}}g_a(z)J^a(z)I\,;~~~g_a(-1/z)=g_a(z)\,,
~~~g_a(\pm i)=0.
\end{eqnarray}
Using the OPE (\ref{eq:OPE_JJb}), we can compute its pure gauge form as
\begin{eqnarray}
&&e^{-\Lambda_LI}\Q e^{\Lambda_LI}
\label{eq:pure_gen_b}\\
&=&-V_L^b(\partial g_a((e^{\cal M}-1){\cal M}^{-1})^a_{~b})I
-{1\over 4}C(\partial g_a\partial g_b
((e^{\cal M}-1){\cal M}^{-1})^a_{~c}g^{cd}
((e^{\cal M}-1){\cal M}^{-1})^b_{~d})I\,,\nonumber
\end{eqnarray}
where we denoted ${\cal M}^a_{~b}(z)={1\over \sqrt{2}}f^{ac}_{~~b}g_c(z)$.
Comparing with the solution (\ref{eq:solution_b}),
we can get a gauge parameter by solving
\begin{eqnarray}
\label{eq:pure_eqb}
 F_b(z)&=&\partial g_a(z)((e^{\cal M}-1){\cal M}^{-1})^a_{~b}(z)\,,
\end{eqnarray}
with respect to $g_a(z)$ for a given $F_a(z)$.
For a solution $g_a(z)$, the new BRS operator $\Q'$ (\ref{eq:QBp_b})
can be rewritten as a similarity transformation from the original one:
\begin{eqnarray}
\label{eq:sim_gen_b}
&&\Q'=e^{-\Lambda}\Q e^{\Lambda},~~~\Lambda=\oint{dz\over 2\pi
 i}{1\over \sqrt{2}}g_a(z)J^a(z).
\end{eqnarray}
In this case, we recover the original SFT action by performing a
field redefinition such as $\Psi''=e^{\Lambda}\Psi'$, which 
implies the effect of a marginal deformation by 
a current $J^a$. Its deformation parameter $g_a(z)$ is related to
the classical solution of string field theory as
(\ref{eq:pure_eqb}).
However, we cannot always obtain a solution $g_a(z)$ to the above
differential equation (\ref{eq:pure_eqb}) because of the boundary
condition $g_a(\pm i)=0$,
which is imposed by a partial integration in computing
eq.~(\ref{eq:pure_gen_b}). 
This situation is just the same as supersymmetric case in 
\S \ref{sec:super_gen}. Namely, consistency condition for 
the pure gauge form (\ref{eq:pure_gen_b}) is given by
(\ref{eq:wilson_non}) because the differential equation
(\ref{eq:pure_eqb}) is the same as (\ref{eq:pure_gen}).
Therefore, there is a possibility that a solution $\Psi_0$
becomes nontrivial if $F_a(z)$ does not satisfy 
eq.~(\ref{eq:wilson_non}).

As an example of ${\cal G}=u(1)^{26}$, 
we take the current $J^{\mu}={i\over \sqrt{2\alpha'}}\partial X^{\mu}$
on the flat background.
In this case, we can identify as
\begin{eqnarray}
&&T(z)=-{1\over 4\alpha'}\partial X^{\mu}\partial X_{\mu}(z)\,,~~~
f^{ab}_{~~c}=0\,, ~~~~\Omega^{\mu\nu}=2\eta^{\mu\nu},~~~c={\rm
 dim}(u(1)^{26})=26,~~~
\end{eqnarray}
using the above notation. 
If some directions are compactified to the torus, 
a solution $\Psi_0$ (\ref{eq:solution_b}) becomes nontrivial
according to a nontrivial Wilson lines: 
$\int_{C_{\rm left}}dzF_{\mu}(z)\ne 0$ \cite{rf:TT1}.
When one direction ($X^{25}$) is 
$S^1$-compactified at the critical radius $R=\sqrt{\alpha'}$, 
we can regard the algebra as ${\cal G}=u(1)^{25}\times su(2)$
and identify as
\begin{eqnarray}
 &&J^1(z)=\sqrt{2}\cos\!\left({X^{25}\over \sqrt{\alpha'}}\right)\!(z),~
~J^2(z)=\sqrt{2}\sin\!\left({X^{25}\over \sqrt{\alpha'}}\right)\!(z),~~
J^3(z)={i\over \sqrt{2\alpha'}}\partial X^{25}(z),~~~~\\
&&T(z)=-{1\over 4\alpha'}\partial X^{25}\partial X^{25}(z)
={1\over 6}:\!(J^1J^1+J^2J^2+J^3J^3)\!:\!(z)\,,\\
&&f^{ab}_{~~c}=\sqrt{2}i\epsilon_{abc}\,,~~~\Omega^{ab}=6\delta^{ab}\,,
~~c={\rm dim}(su(2))-3\cdot 4/6=1\,,
\end{eqnarray}
in the $su(2)$ sector. The corresponding solution $\Psi_0$ was
investigated in refs.~\cite{rf:TT1, rf:KTZ, Kluson_02}.

%%%%%%%%%%%%%%%%%%%%%%%%%%%%%%%%%%%%%%%%%%%%%%%%%%%%%%%%%%%%%%
%\newpage

%%%%%%%%%%%%%%%%%%%%%%%%%%%%%%%%%%%%%%%%%%%%%%%%%%%%%
\end{document}